\documentclass[lettersize,journal]{IEEEtran}
\usepackage{amsmath,amsfonts}
\usepackage{amssymb}
\usepackage{amsthm}

\usepackage{subfigure}
\usepackage{booktabs, tabularx}
\usepackage{algorithmic}
\usepackage{algorithm}
\usepackage{enumitem}
\usepackage{bm}
\usepackage{array}
\usepackage[caption=false,font=normalsize,labelfont=sf,textfont=sf]{subfig}
\usepackage{textcomp}
\usepackage{stfloats}
\usepackage{url}
\usepackage{verbatim}
\usepackage{graphicx}
\usepackage{cite}
\usepackage{xcolor}
\usepackage[normalem]{ulem}

\usepackage{xcolor}
\definecolor{myblue}{RGB}{16,110,173}

\newcommand{\add}[1]{#1}
\newcommand{\del}[1]{}

\begin{document}

\title{Stackelberg Game with Multi-Agent Transformer for ISRS-Aware Resource Re-optimization in Multi-Band Optical Networks}

\author{Jiaxin Liu,
	Rentao Gu,~\IEEEmembership{Senior Member,~IEEE,}
	Miao Zhu,
	Zeguang Zou,
	Hui Li,~\IEEEmembership{Member,~IEEE,}
	Yuefeng Ji,~\IEEEmembership{Senior Member,~IEEE}
\thanks{This work is supported by National Natural Science Foundation of China (U25B2013), P. R. China. (Corresponding author: Rentao Gu)}
\thanks{Jiaxin Liu, Rentao Gu, Miao Zhu, Zeguang Zou, Hui Li, and Yuefeng Ji are with State Key Lab of Information Photonics and Optical Communications, Beijing University of Posts and Telecommunications, Beijing 100876, China (e-mail: jiaxinliu@bupt.edu.cn; rentaogu@bupt.edu.cn; zmiao142857@bupt.edu.cn; zouzeguang@bupt.edu.cn; lihui@bupt.edu.cn; jyf@bupt.edu.cn).}
}


\maketitle

\begin{abstract}
The rapid expansion of generative AI is driving computing infrastructures toward interconnected AI data centers. Multi-band optical networks have become the backbone of such connectivity. To accommodate increasingly dynamic traffic demands, these networks require continual resource re-optimization to improve resource utilization and reduce service blocking. However, multi-band transmission amplifies inter-channel stimulated Raman scattering (ISRS), inducing nonlinear coupling among services. This coupling complicates re-optimization, as local adjustments perturb the global power profile, creating interdependent optimization decisions across services. Moreover, the nonlinear interactions vary with the subset of services selected for re-optimization, and inappropriate selection may \del{destabilize the process} \add{reduce re-optimization effectiveness} or diminish achievable benefits. To address these challenges, we propose a Stackelberg Game with Multi-Agent Transformer (SG-MAT) approach for ISRS-aware resource re-optimization. SG-MAT employs a leader–follower structure, in which the MAT-based follower models nonlinear coupling and generates coordinated channel reallocations, feeding the utilization improvements to the leader. Using this feedback, the leader identifies critical service subsets, approximating a Stackelberg equilibrium to guide global decisions and enabling \del{stable} \add{consistent} and efficient re-optimization. Simulation results show that SG-MAT achieves the highest utilization improvement and reduces the blocking ratio by approximately 30\% under moderate traffic loads, with computational complexity comparable to benchmarks.
\end{abstract}

\begin{IEEEkeywords}
Multi-band optical networks, inter-channel stimulated Raman scattering, nonlinear coupling, multi-agent Transformer, Stackelberg equilibrium.
\end{IEEEkeywords}

\section{Introduction}

\IEEEPARstart{D}{riven} by the rapid proliferation of generative AI \cite{Kun Qian, C. -Y. Liu, J. Zhu}, AI data center (AIDC) infrastructures are evolving from localized, single-site systems to distributed and collaboratively operated architectures \cite{N. Jones, X. Xie}. This transformation imposes stringent requirements on network performance, particularly in throughput, latency, and reliability performance. Optical networks, with their high bandwidth, ultra-low latency, and superior energy efficiency, have become the dominant architecture for inter-AIDC networks \cite{Z. Chen, Y. Pang, J. Zhu, F. Li}. As inter-AIDC traffic continues to surge, multi-band optical transmission has become essential to support this growing traffic and overcome the capacity constraints of single-band systems.

\begin{figure}[tp]
	\centering
        \includegraphics[width=0.48\textwidth]{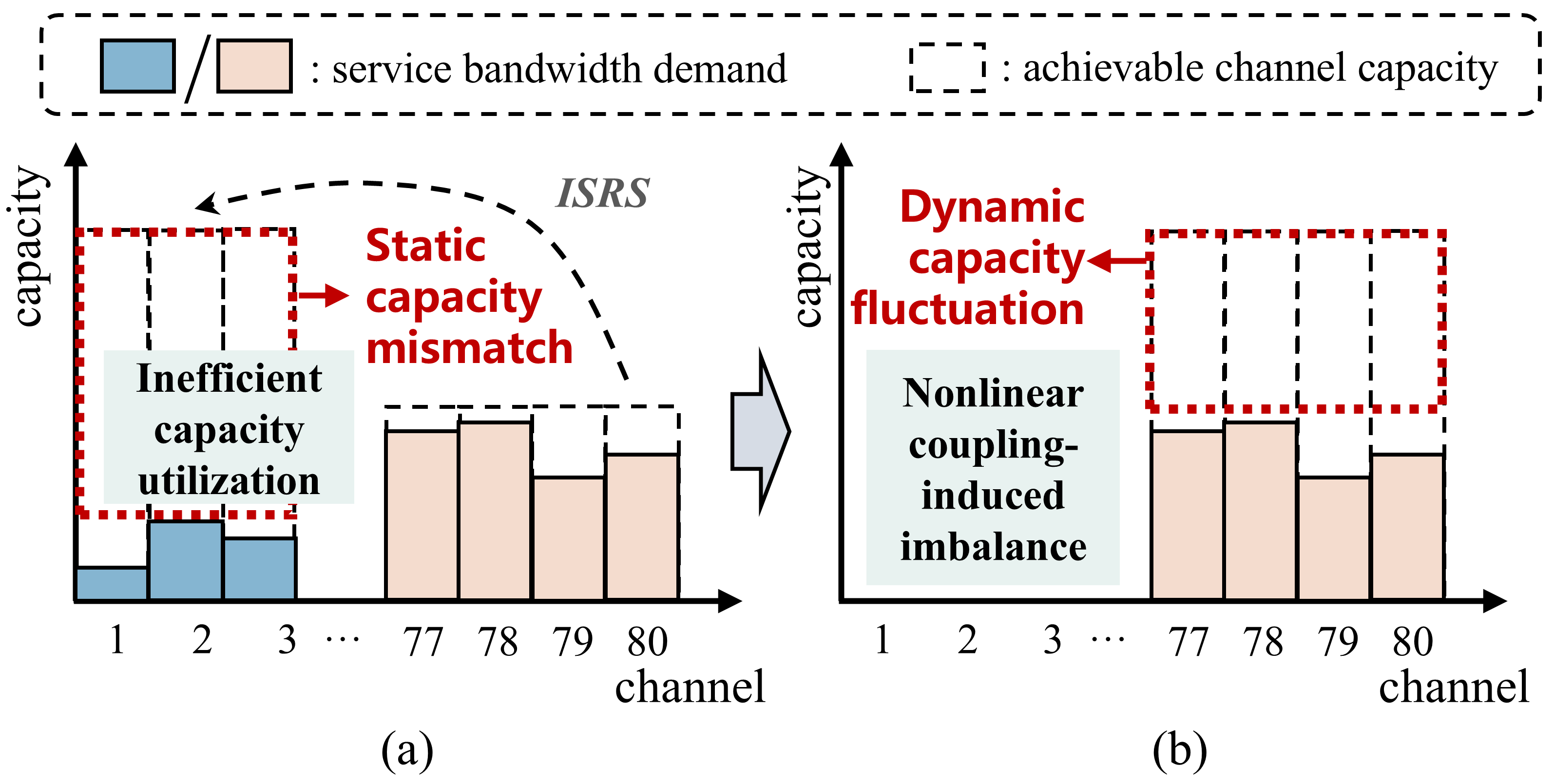}
	\caption{Illustration of resource imbalance in MB-WDM networks. (a) Static capacity mismatch under heterogeneous workloads. (b) Dynamic capacity fluctuation and resource imbalance under ISRS.}
	\label{fig:background}
\end{figure}

Nevertheless, the widespread deployment of multi-band wavelength-division multiplexed (MB-WDM) optical networks introduces new challenges arising from traffic behaviors and physical-layer effects. The inherently heterogeneous AIDC workloads, characterized by frequent arrivals and releases with diverse bitrates and durations, continuously perturb the link power spectrum and destabilize network operation. Meanwhile, inter-channel stimulated Raman scattering (ISRS) at the physical layer induces cross-channel power transfer, which dynamically redistributes signal power as services are established, released, or reconfigured \cite{S. Li, C. Zhang}. These effects jointly cause variations in generalized signal-to-noise ratio (GSNR) and channel capacity. Under a WDM architecture with limited spectral flexibility, such perturbations aggravate the mismatch between available channel capacity and service bandwidth demand, as illustrated in Fig. \ref{fig:background}. These effects collectively lead to resource imbalance, service blocking, and throughput degradation in MB-WDM optical networks.

To address the aforementioned challenges, real-time resource re-optimization, achieved through coordinated service-channel reassignment, is essential for improving resource utilization under dynamic and nonlinear conditions. However, such multi-service reconfiguration introduces intricate inter-service dependencies. Adjusting one service perturbs the ISRS-induced power distribution along the spectrum, which in turn alters the GSNR and feasible resource allocation of other co-propagating services. Consequently, the global performance gain becomes inherently non-additive, as multi-service reconfigurations produce system-level effects that cannot be inferred from individual adjustments due to their nonlinear interactions. Moreover, distinct service subsets give rise to different nonlinear interaction dynamics, producing global outcomes that can vary substantially with the particular subset chosen. Existing studies often overlook these nonlinear interdependencies among services \cite{S. S. Johari, B. Bao, E. Etezadi}. As a result, inappropriate subset selection or suboptimal reassignment decisions often reduce network-wide efficiency.

\del{In addition, frequent service arrivals, releases, and reconfigurations introduce temporal variability to the network state, making the re-optimization problem inherently non-stationary, as the state distribution shifts over time and renders past policies unreliable. Traditional heuristics and offline-trained models lack mechanisms to incorporate real-time feedback. Consequently, their policies exhibit poor adaptability and weak generalization under dynamic network environments~\cite{Y. Dang, A. Ben-Ameur}. Although online reinforcement learning (ORL) enables adaptive control through real-time policy refinement~\cite{S. Emara, J. Li}, policy oscillation and convergence instability hinder its reliability in optical networks. These limitations call for a robust optimization framework that ensures stable and responsive adaptation under time-varying nonlinear dynamics.}

\add{In addition, frequent service arrivals, releases, and reconfigurations continually reshape the instantaneous network state, so blocking events differ in spectrum occupancy, ISRS-induced power profiles, GSNR margins, and candidate services for re-optimization. These event-level variations limit fixed rules and offline-trained policies under changing network conditions~\cite{Y. Dang, A. Ben-Ameur}. Although online reinforcement learning (ORL) improves adaptability through environmental interaction~\cite{S. Emara, J. Li}, the combinatorial decision space and nonlinear service coupling can still hinder exploration and make re-optimization gains less consistent. This calls for a structured coordination mechanism that models service-subset selection and ISRS-aware channel reassignment as coupled decisions under the current network state.}

We propose an adaptive, ISRS-aware resource re-optimization mechanism based on Stackelberg Game with Multi-Agent Transformer (SG-MAT) to \del{handle both nonlinear coupling and non-stationary dynamics} \add{address nonlinear inter-service coupling and event-level state variations in multi-band optical networks}. SG-MAT adopts a leader–follower structure, where the MAT-based follower models ISRS-induced inter-service coupling and provides utility feedback to the leader, while the leader selects the critical service subset that approximates Stackelberg equilibrium (SE). The main contributions of this work are summarized as follows.

\begin{enumerate}[label=(\arabic*)]
	\item We develop an ISRS-aware resource re-optimization mechanism that adaptively adjusts service channels by explicitly modeling ISRS-induced power migration and its impact on GSNR. The mechanism leverages constructive ISRS effects to improve effective bandwidth utilization and network capacity, alleviates bandwidth mismatch and substantially reduces blocking.
	\item We introduce MAT that integrates the Multi-Agent Advantage Decomposition Theorem with attention-driven interaction modeling. This architecture captures ISRS-induced cross-service interference propagation, identifies interdependencies among co-propagating services, and enables joint multi-service reconfiguration. MAT eliminates the sequential-decision interference and significantly enhances global scheduling efficiency.
	\item We formulate the re-optimization problem as a leader–follower Stackelberg game and develop SG-MAT. Through iterative interaction between the leader and follower, where the follower provides utilities of candidate reallocations based on MAT and the leader updates the service subset accordingly, SG-MAT progressively approximates SE. This iterative equilibrium-driven process enables \del{stable and real-time re-optimization in non-stationary multi-band optical networks} \add{consistent and efficient re-optimization decisions under dynamic network states and nonlinear ISRS-induced coupling}.

\end{enumerate}

The remainder of this paper is organized as follows. Section II reviews related research. Section III establishes the overall system model and formulates the Stackelberg game. Section IV introduces MAT strategy for modeling nonlinear multi-service coupling and deriving best coordinated follower responses, while Section V establishes the leader optimization based on backward induction to approximate SE. Section VI provides the experimental evaluation and performance analysis. Finally, Section VII summarizes the paper.

\section{Related Works}
\subsection{Resource Re-optimization under Nonlinear Physical Impairments}
Resource re-optimization has long been used to improve resource utilization and reduce service blocking in optical networks. Existing studies have addressed this problem from fragmentation prediction, modeling, and mitigation perspectives. Chatterjee et al.~\cite{B. C. Chatterjee} proposed traffic-prediction-driven proactive fragmentation management, while Ahmed et al.~\cite{I. Ahmed} characterized the probabilistic relationship between fragmentation and blocking. To handle heterogeneous and time-varying demands, Bao et al.~\cite{B. Bao} developed a time-aware fragmentation-sensitive allocation algorithm, Zou et al.~\cite{Z. Zou} extended fragmentation mitigation to the joint time--frequency--space domain, and Rezagholizadeh et al.~\cite{E. Rezagholizadeh} proposed an auxiliary-graph co-optimization framework for multi-layer optical networks.

The emergence of multi-band optical systems introduces additional challenges arising from non-negligible nonlinear impairments that couple across frequencies. Teng et al.~\cite{Y. Teng} developed an ISRS-aware allocation strategy for C+L-band systems by jointly considering spectral and temporal characteristics. Yang et al.~\cite{Z. Yang} proposed a hypergraph-based ISRS-aware mapping method to capture multi-path power coupling, while Ravipudi et al.~\cite{J. L. Ravipudi} jointly considered crosstalk, ASE noise, and nonlinear impairments for energy-efficient and fragmentation-aware provisioning. Ren et al.~\cite{D. Ren} further incorporated real-time optical performance monitoring into the allocation loop, enabling closed-loop adaptive control.

In parallel, learning-driven resource scheduling has gained increasing traction as a means to address the complexity of optical resource management. Moore et al.~\cite{F. Moore} exploited statistical traffic distributions for probabilistic path--spectrum assignment, and Asiri~\cite{A. Asiri} applied deep reinforcement learning to QoT-aware routing and spectrum assignment. These studies represent important steps toward data-driven and adaptive control paradigms in optical networking.

However, most existing approaches fail to explicitly capture the inter-service dependencies induced by ISRS. As a result, the mutual influence among co-propagating services is insufficiently captured, limiting coordinated reconfiguration under nonlinear physical impairments. This motivates ISRS-aware re-optimization frameworks that can model cross-service interactions and support performance-aware resource coordination under nonlinear and time-varying conditions.

\subsection{Adaptive Resource Management \del{in Non-stationary Environments} \add{under Dynamic Network Conditions}}

In dynamic optical networks, resource allocation and reconfiguration are affected by time-varying traffic demands and physical-layer impairments.
\del{The stochastic and non-stationary nature of these environments significantly complicates decision-making, as network states evolve continuously and historical policies quickly become suboptimal.} \add{Such state evolution makes reconfiguration decisions depend strongly on the current spectrum occupancy and physical-layer conditions of each blocking event.} Early studies relied mainly on heuristic and rule-based defragmentation. Comellas et al. \cite{Comellas} devised several proactive defragmentation strategies for elastic optical networks (EONs), achieving near-optimal performance through priority-based reorganization rules. While effective in specific conditions, such static schemes \del{inherently lack adaptability to real-time variations and nonlinear interference dynamics} \add{remain limited in adapting reconfiguration decisions to real-time spectrum occupancy and physical-layer conditions}.

To overcome these limitations, learning-based techniques have been introduced to improve data-driven adaptation. Etezadi et al. \cite{E. Etezadi} employed DRL to determine the optimal re-optimization timing in dynamically evolving environments, while Johari et al. \cite{S. S. Johari} showed that DRL still faces scalability, convergence, and interpretability issues in high-dimensional settings. Moreover, offline-trained models \del{inherently lack the flexibility to continuously adapt to evolving network states in highly dynamic optical environments} \add{may suffer degraded performance when the operating state differs from the training conditions}. Consequently, such approaches are not the primary focus of this work.

Motivated by these challenges, recent studies have explored \del{online reinforcement learning (ORL)} \add{ORL} for policy adaptation. Emara et al. \cite{S. Emara} proposed a Pareto-based ORL (PORL) framework incorporating adaptive multi-objective weighting, thereby facilitating on-policy learning in non-stationary and conflicting objective spaces. Li et al. \cite{J. Li} further developed a self-distillation ORL (SDORL) scheme that periodically transfers accumulated knowledge across training phases and applies $L_2$-norm regularization to mitigate overfitting and early-experience bias, yielding notable improvements in convergence stability and policy generalization. \add{Ayepah-Mensah et al. \cite{Ayepah} proposed a federated policy distillation framework that enhances learning stability and knowledge transfer under heterogeneous, non-i.i.d., and non-stationary conditions.}
\del{However, ORL-based approaches in high-dimensional optical networks often lead to slow convergence and high gradient variance. Moreover, the lack of structural constraints in purely data-driven optimization can result in policy oscillation or unstable control feedback under dynamic conditions.} \add{However, ORL remains challenging for multi-band optical networks, where the combinatorial decision space and nonlinear service coupling can hinder efficient exploration and result in variable re-optimization gains.}

\add{Game-theoretic methods provide a complementary way to structure adaptive resource management. Existing studies have applied distributed Stackelberg games to leader--follower resource coordination~\cite{Zhang}, Stackelberg evolutionary games to dynamic strategy adaptation~\cite{Shen}, and non-cooperative or attacker--defender games to robust resource control~\cite{Lian}. These works show that utility modeling, equilibrium analysis, and interaction mechanisms can support interpretable and dynamically adaptive decision-making.}

\add{ORL and game-theoretic coordination thus address adaptive resource management from complementary perspectives. The former improves learning stability and generalization through knowledge transfer and regularization, whereas the latter structures decision-making through hierarchical, competitive, or cooperative interaction modeling.}

\section{System Model and Stackelberg Game Formulation}
In this section, we first present the underlying physical-layer model of the considered multi-band optical network and the associated service model, and then formulate the resource re-optimization problem as a Stackelberg leader–follower game.

\subsection{Network Model and Physical Layer Model with ISRS}
We consider a C+L-band WDM optical network, modeled as a directed graph
\begin{align}
    G=(V,E),
\end{align}
where $V$ and $E$ denote the sets of nodes and fiber links, respectively. Each link $e_{sr}\in E$ between nodes $n_s$ and $n_r$ consists of one or more fiber spans, each with length no greater than $L_{\max}$, and supports $W$ parallel WDM channels with per-channel bandwidth $B$. C- and L-band channels are multiplexed and co-propagate along each span, experiencing ISRS-induced power evolution during propagation. At the span output, signals are demultiplexed and fully loss-compensated by erbium-doped fiber amplifiers (EDFAs)~\cite{N. Guo}. To mitigate inter-band power tilt induced by ISRS, a tunable gain-flattening filter is cascaded after each EDFA \cite{M. Mehrabi}. Reconfigurable add-drop multiplexers (ROADMs) provide reconfigurable multi-band add/drop and switching at nodes, with the insertion loss compensated by terminal amplification \cite{Z. Yang, J. Kundrát}.

Let $S=\{s_1,s_2,\ldots\}$ denote the set of service requests. Each request $s_i$ is specified by
\begin{align}
    (s_i^{\text{src}},s_i^{\text{dst}},b_i,t_i^{\text{arr}}, t_i^{\text{hold}}),
\end{align}
where $s_i^{\text{src}}$ and $s_i^{\text{dst}}$ are the source and destination nodes, $b_i$ is the bitrate modeled as a random variable to reflect AI-task heterogeneity, $t_i^{\text{arr}}$ is the arrival time, and $t_i^{\text{hold}}$ is the holding time. Arrivals follow a Poisson process with rate $\lambda$, holding times are exponentially distributed with mean $1/\mu$, and the traffic load is $\lambda/\mu$ Erlang.

For a candidate route, the QoT is evaluated and the highest admissible modulation order satisfying a target bit-error rate (BER) is selected to maximize spectral efficiency~\cite{N. Guo}. Since nonlinear effects are non-negligible in multi-band operation, we adopt GSNR as a conservative QoT metric. For channel $j$, the GSNR is defined as
\begin{align}
    \mathrm{GSNR}_j \;\triangleq\; \frac{P_{\text{s},j}}{P_{\text{ASE},j}+P_{\text{NLI},j}},
    \label{eq:GSNR_cal}
\end{align}
\noindent where $P_{\text{s},j}$ is the received signal power, $P_{\text{ASE},j}$ is the ASE noise power, and $P_{\text{NLI},j}$ denotes the nonlinear-interference (NLI) power, all evaluated at the span output.

We adopt the generalized stimulated Raman scattering model \cite{C. Lasagni} to model the received signal power at channel $j$, approximated by
\begin{align}
	P_{\text{s},j} \approx  P_{\text{l},j} \cdot \frac{e^{-C_rL_{\text{eff}}(z)r(f_j)-\alpha_jz}P_t}{\sum\limits_{m=1}^{W} P_{\text{l},m}e^{-C_rL_{\text{eff}}(z)r(f_m)}},
\end{align}
\noindent where $P_{\text{l},j}$ is the launch power of channel $j$, $C_r$ is the Raman gain slope, $ L_{\text{eff}}(z)=(1-e^{-\alpha_j z})/\alpha_j$ is the effective span length, $\alpha_j$ is the fiber attenuation at channel $j$, $z$ denotes the transmission distance of the last span, $r(\cdot)$ is the normalized Raman-gain profile, $f_j$ is the central frequency of channel $j$, and $P_t=\sum_{m=1}^{W}P_{\text{l},m}$ is the total launched power. 

The ASE power of channel $j$ accumulated over $N$ spans is
\begin{align}
	P_{\text{ASE},j} = \sum\limits_{s=1}^{N} 2n_{\text{sp}}hf_jB(G_s-1),
\end{align}
\noindent where $n_{\text{sp}}$ is the spontaneous emission factor, $h$ is Planck’s constant, $B$ is the channel bandwidth, and $G_s$ is the $s$-th span amplifier gain. 

Following the Gaussian Noise approximation~\cite{D. Semrau}, the NLI power of channel $j$ accumulated over $N$ spans is
\begin{align}
	P_{\text{NLI},j} = \eta(f_j) P_{\text{l},j}^3,
\end{align}

\noindent with the NLI coefficient $\eta(f_j)$ given by
\begin{align}
	\eta(f_j) \approx \sum\limits_{s=1}^{N} \left[\frac{P_{\text{s},j}}{P_{\text{l},j}}\right]^3 \cdot \big[\eta_{\text{SPM}}(f_j) N^{\varepsilon} + \eta_{\text{XPM}}(f_j)\big],
\end{align}

\noindent where $\varepsilon$ is the coherence factor. The self-phase modulation (SPM) and cross-phase modulation (XPM) contributions are \del{detailed in (\ref{eq:SPM}) and (\ref{eq:XPM}).} \add{given by}

{
\setlength{\abovedisplayskip}{2pt}
\setlength{\belowdisplayskip}{2pt}
\begin{align}
\eta_{\mathrm{SPM}}(f_j) \approx&
\frac{4\gamma^2\pi}
{9B^2\phi_j\bar{\alpha}(2\alpha_j+\bar{\alpha})}
\bigg[
\frac{T_j-\alpha_j^2}{\alpha_j}
\operatorname{asinh}\!\left(\frac{\phi_j B^2}{\pi\alpha_j}\right) \notag\\
&\quad+
\frac{A_j^2-T_j}{A_j}
\operatorname{asinh}\!\left(\frac{\phi_j B^2}{\pi A_j}\right)
\bigg],
\label{eq:SPM}
\end{align}
}
{
\setlength{\abovedisplayskip}{2pt}
\setlength{\belowdisplayskip}{2pt}
\begin{equation}
\begin{aligned}
&\eta_{\text{XPM}}(f_j) \approx {}
\frac{32}{27}
\sum\limits_{k=1,\;k\neq j}^{W}
\left( \frac{P_{\text{s},k}}{P_{\text{l},j}} \right)^2
\frac{\gamma^2}{B\phi_{j,k}\bar{\alpha}(2\alpha_k+ \bar{\alpha})} \\
& \cdot
\left[
\frac{T_k - \alpha_k^2}{\alpha_k}
\operatorname{atan}\!\left( \frac{\phi_{j,k} B}{\alpha_k} \right)
+
\frac{A_k^2 - T_k}{A_k}
\operatorname{atan}\!\left( \frac{\phi_{j,k} B}{A_k} \right)
\right],
\end{aligned}
\label{eq:XPM}
\end{equation}
}

\noindent \add{where} $\gamma$ is the fiber nonlinear coefficient, $\phi_j=\tfrac{3}{2}\pi^2(\beta_2+2\pi\beta_3 f_j)$, $\beta_2$ is the group velocity dispersion, $\beta_3$ is the dispersion slope, $\bar{\alpha}$ is the effective attenuation, $T_j=(\alpha_j+\bar{\alpha}-P_t C_r f_j)^2$, $A_j=\alpha_j+\bar{\alpha}$, and $\phi_{j,k}=2\pi^2(f_k-f_j)\big[\beta_2+\pi\beta_3(f_j+f_k)\big]$.

\subsection{Stackelberg Game Formulation of the ISRS-Aware Resource Re-optimization Problem}
\label{sec:Stackelberg Game Formulation}
To mitigate service blocking caused by inefficient resource utilization in multi-band optical networks, we develop a reactive ISRS-aware re-optimization scheme. Upon each blocking event, a subset of ongoing services is selected and reassigned to alternative channels to improve global resource utilization and reduce subsequent blocking.
\add{We consider a single-administrative-domain scenario managed by a logically centralized control unit, such as a Software-Defined Networking (SDN) controller~\cite{T. Xia}. This architecture is consistent with existing optical-network control frameworks, where the controller maintains network topology, spectrum occupancy, service states, and physical-layer monitoring information, and coordinates device reconfiguration through southbound interfaces.}
\del{This procedure is modeled as a Stackelberg leader–follower game, where the leader selects the service subset and the follower performs the corresponding channel reassignment.}
\add{Within this control architecture, SG-MAT is deployed as an application-layer re-optimization module and follows a Stackelberg leader--follower formulation, where the leader performs service-subset selection and the follower performs the corresponding channel reassignment.}
The main notations \add{used in the Stackelberg formulation} are summarized in Table~\ref{tab:notations-sets}, and the Stackelberg model is decomposed into the leader and follower subproblems as follows.

\begin{table}[t]
	\centering
	\caption{\add{Summary of Main Notations in the Stackelberg Formulation}}
	\label{tab:notations-sets}
	\footnotesize
	\begin{tabularx}{\columnwidth}{lX}
		\toprule
		\textbf{Notation} & \textbf{Description} \\
		\midrule
		$S_p$ & Candidate service set within the permissible re-optimization scope. \\
		$s_i$ & The $i$-th candidate service in $S_p$. \\
		$I$ & Number of candidate services, $I=|S_p|$. \\
		$\mathbf{x}$ & Leader decision vector, $\mathbf{x}=(x_1,\ldots,x_I)\in\{0,1\}^{I}$. \\
		$x_i$ & Binary indicator; $x_i=1$ if service $s_i$ is selected for re-optimization, and $x_i=0$ otherwise. \\
		$\mathcal{S}(\mathbf{x})$ & Set of services selected by the leader, $\mathcal{S}(\mathbf{x})=\{s_i\mid x_i=1\}$. \\
		$\mathcal{F}$ & Set of available central frequencies for channel reassignment, $\mathcal{F}=\{\nu_1,\nu_2,\ldots,\nu_W\}$. \\
		$\mathcal{W}$ & Set of channel indices, $\mathcal{W}=\{1,2,\ldots,W\}$, where index $w$ corresponds to frequency $\nu_w$. \\
		$\mathbf{f}$ & Channel configuration of $S_p$ after re-optimization, $\mathbf{f}=(f_1,\ldots,f_I)$. \\
		$f_i$ & Central frequency assigned to service $s_i$ after re-optimization. \\
		$f_i^{\mathrm{ori}}$ & Original central frequency of service $s_i$ before re-optimization. \\
		$y_{s_i,w}$ & Binary channel-assignment variable; $y_{s_i,w}=1$ if service $s_i$ is assigned to frequency $\nu_w$, and $0$ otherwise. \\
		$y_{s_i,w}^{\mathrm{ori}}$ & Original channel-assignment indicator of service $s_i$ before re-optimization. \\
		$a_{e,w}$ & Binary occupancy indicator; $a_{e,w}=1$ if frequency $\nu_w$ on link $e$ is occupied by a service outside the current candidate set, and $0$ otherwise. \\
		$\mathcal{P}_{s_i}$ & Set of fiber links on the route of service $s_i$, $\mathcal{P}_{s_i}\subseteq E$. \\
		$K_i$ & Path length of service $s_i$ in terms of link count, $K_i=|\mathcal{P}_{s_i}|$. \\
		$b_i$ & Bitrate demand of service $s_i$. \\
		$C_i^e(\mathbf{f})$ & Achievable capacity of service $s_i$ on link $e$ under channel configuration $\mathbf{f}$. \\
		$C_i(\mathbf{f})$ & End-to-end achievable capacity of service $s_i$, $C_i(\mathbf{f})=\min_{e\in\mathcal{P}_{s_i}}C_i^e(\mathbf{f})$. \\
		$\mathrm{GSNR}_i^e(\mathbf{f})$ & GSNR of service $s_i$ on link $e$ under channel configuration $\mathbf{f}$. \\
		$f_{\mathrm{mod}}(\cdot)$ & Mapping from GSNR to achievable modulation-dependent capacity. \\
		$\Delta\mathrm{Util}_i(\mathbf{f})$ & Average per-link bandwidth-utilization gain of selected service $s_i$. \\
		$\Delta\mathrm{GSNR}_j(\mathbf{f})$ & Average GSNR improvement of unselected service $s_j$. \\
		$U_F(\mathbf{x},\mathbf{f})$ & Follower utility under leader decision $\mathbf{x}$ and channel configuration $\mathbf{f}$. \\
		$U_L(\mathbf{x})$ & Leader utility under service-selection decision $\mathbf{x}$. \\
		$\mathbf{f}^*(\mathbf{x})$ & Follower best response under leader decision $\mathbf{x}$. \\
		$p_i$ & Reconfiguration cost weight associated with service $s_i$. \\
		$k$ & Weight balancing utilization gain and GSNR improvement in the follower utility. \\
		$\overline{|\mathcal{P}|}$ & Average path length of affected services in terms of link count. \\
		\bottomrule
	\end{tabularx}
\end{table}

\subsubsection{Leader Formulation}
The leader determines the subset of services to be re-optimized. Let $S_p=\{s_1,\dots,s_I\}$ denote the set of services within the permissible re-optimization scope, where $I=|S_p|$. The leader decision vector is defined as

\begin{align}
	\mathbf{x}=(x_1,\ldots,x_I)\in\mathcal{X} \triangleq \{0,1\}^{I},
\end{align}
where $x_i=1$ indicates that service $s_i$ is selected for re-optimization. The subset of services selected for re-optimization is then given by
\begin{align}
	\mathcal{S}(\mathbf{x}) = \{ s_i\mid x_i=1\}.
\end{align}

The leader utility $U_L(\mathbf{x})$ quantifies the network utilization and GSNR gain induced by re-optimization minus the re-optimization cost, defined as
\begin{align}
	U_L(\mathbf{x})=U_F(\mathbf{x},\mathbf{f}^*(\mathbf{x}))-\sum_{i=1}^{I}{p_ix_i},
	\label{eq:U_L}
\end{align}
where $U_F(\mathbf{x},\cdot)$ is the follower utility, $p_i$ denotes the reconfiguration cost associated with service $s_i$ (with $p_i x_i=0$ when $x_i=0$), and $\mathbf{f}^*(\mathbf{x})$ denotes the follower’s best response given $\mathbf{x}$. The cost weights $p_i$ adapt online according to network load and operational priorities.

Accordingly, the leader problem is expressed as
\begin{align}
	\text{(P1)}\quad 
	& \max_{\mathbf{x}\in\{0,1\}^I} U_L(\mathbf{x}) \\
	\text{s.t.}\quad
	& \mathbf{f}^*(\mathbf{x})\in\arg\max_{\mathbf{f}} U_F(\mathbf{x},\mathbf{f}). 
\end{align}

\subsubsection{Follower Formulation}
Given the leader-determined set $\mathcal{S}(\mathbf{x})$ of services to be adjusted, the follower performs channel reassignment so as to maximize the resulting network-wide performance gain. Let
\begin{align}
    \mathbf{f}=\{f_1,f_2,\ldots,f_I\} \in \mathcal{F}^I
\end{align}
\del{denote the follower decision vector, where $f_i$ is the central frequency of post-reoptimization channel of service $s_i \in S_p$.}
\add{denote the channel configuration of $S_p$ after re-optimization, where $f_i$ is the central frequency assigned to service $s_i$.}
For services not selected by the leader, the original channel is retained, i.e., $f_i=f_i^{\mathrm{ori}}$ for $s_i\in S_p\setminus\mathcal{S}(\mathbf{x})$.

The follower utility aggregates the average utilization gain of re-optimized services and the average GSNR improvement of other services:
\begin{align}
	&U_F(\mathbf{x},\mathbf{f})=\frac{1}{\max\{1,|\mathcal S(\mathbf x)|\}}\sum_{s_i\in\mathcal{S}(\mathbf{x})}{\Delta{\mathrm{Util}}_i(\mathbf{f})} \notag
	\\
	& +k\cdot\frac{1}{\max\{1,|S_p|-|\mathcal S(\mathbf x)|\}}\sum_{s_j\in S_p \setminus \mathcal{S}(\mathbf{x})}{\Delta{\mathrm{GSNR}}_j(\mathbf{f})},
	\label{eq:U_F}
\end{align}
where $k$ balances the contributions of utilization and GSNR improvement. The utilization gain of service $s_i$ is defined as the average per-link bandwidth utilization improvement:
\begin{align}
	\Delta{\mathrm{Util}}_i(\mathbf{f})
	=
	\frac{1}{K_i}
	\sum_{e\in\mathcal{P}_{s_i}}
	\!\left(
	\frac{b_i}{C_{i}^e(\mathbf{f})}
	-
	\frac{b_i}{C_{i}^e(\mathbf{f}^{\mathrm{ori}})}
	\right),
\end{align}
where $C_{i}^e(\mathbf{f})$ denotes the capacity under channel assignments $\mathbf{f}$, and $\mathbf{f}^{\text{ori}}$ is the original frequencies:
\begin{align}
	C_{i}^e(\mathbf f)
	= f_{\mathrm{mod}}(\mathrm{GSNR}_{i}^e(\mathbf f)),
\end{align}
where $\mathrm{GSNR}_{i}^e(\mathbf{f})$ is the GSNR of $s_i$ on link $e$ under $\mathbf{f}$, calculated by (\ref{eq:GSNR_cal}), and $f_{\text{mod}}(\cdot)$ maps the GSNR value to achievable modulation-dependent capacity.

The GSNR improvement of service $s_j$ is
\begin{align}
	\Delta{\mathrm{GSNR}}_j(\mathbf{f})=\frac{1}{K_j}\sum_{e\in\mathcal{P}_{s_j}}{(\mathrm{GSNR}_{j}^e(\mathbf{f})-\mathrm{GSNR}_{j}^e(\mathbf{f}^{\mathrm{ori}}))},
\end{align}
which discourages the optimizer from improving utilization at the cost of degrading overall transmission capacity.

The follower problem can thus be written as 
\begin{align}
	\text{(P2)}\quad
	& \max_{\mathbf f}\ U_F(\mathbf x,\mathbf f) \\[2pt]
	\text{s.t.}\quad
	& y_{s_i,w}=y_{s_i,w}^{\mathrm{ori}}, 
	\quad \forall s_i\in S_p\setminus\mathcal{S}(\mathbf{x}),\ \forall w\in\mathcal W, \label{eq:P2-0}\\
        & \sum_{w \in\mathcal W} y_{s_i,w }=1, \quad \forall s_i\in S_p, \label{eq:P2-1}\\
	& \sum_{i:\, e\in\mathcal P_{s_i}} y_{s_i,w} + a_{e,w} \le 1, \quad \forall e\in E,\ \forall w\in\mathcal W, \label{eq:P2-2}\\
	& b_i \le C_{i}(\mathbf f) = \min_{e\in \mathcal P_{s_i}} C_i^e(\mathbf f), \quad \forall s_i\in S_p. \label{eq:P2-3}
\end{align}

The constraints in (P2) enforce both conflict-free channel assignment and QoT-feasible capacity allocation. (\ref{eq:P2-0}) keeps the original channel unchanged for services not selected by the leader, so that only services in $\mathcal{S}(\mathbf{x})$ are reassigned. (\ref{eq:P2-1}) enforces per-service channel exclusivity, while (\ref{eq:P2-2}) ensures link-level orthogonality by prohibiting concurrent use of the same channel on a given fiber link. (\ref{eq:P2-3}) ensures QoT feasibility by restricting each service’s bandwidth demand within the GSNR-determined modulation capacity, thereby avoiding transmission errors.

\vspace{0.3em}
\textit{Complexity Analysis:}
The routing and wavelength assignment problem in multi-band optical networks is NP-hard~\cite{R. Matzner}, and the cooperative re-optimization of multiple services further amplifies the combinatorial complexity. In the proposed Stackelberg framework, the leader layer performs a binary selection over \(S_p\) with a search space of \(\mathcal{O}(2^{|S_p|})\). For a given leader decision \(\mathbf{x}\), let \(|\mathcal{S}(\mathbf{x})|\) denote the number of services selected by the leader. The follower layer must assign one of the \(W\) available channels to each of the \(|\mathcal{S}(\mathbf{x})|\) services, yielding a discrete assignment space of size \(W^{|\mathcal{S}(\mathbf{x})|}\). Evaluating one candidate assignment requires checking QoT on all links traversed by the affected services. Let \(\overline{|\mathcal{P}|}\) denote the average path length in terms of link count. Accordingly, the computational complexity of the proposed Stackelberg re-optimization framework is
\begin{align}
	\mathcal{O}\!\left(2^{|S_p|} \cdot W^{|\mathcal{S}(\mathbf{x})|} \cdot \overline{|\mathcal{P}|}\right).
\end{align}

\section{Follower Response under Nonlinear Coupling: A Multi-Agent Transformer Strategy}
\label{sec:MAT}
\begin{figure}[tp]
	\centering
        \includegraphics[width=0.48\textwidth]{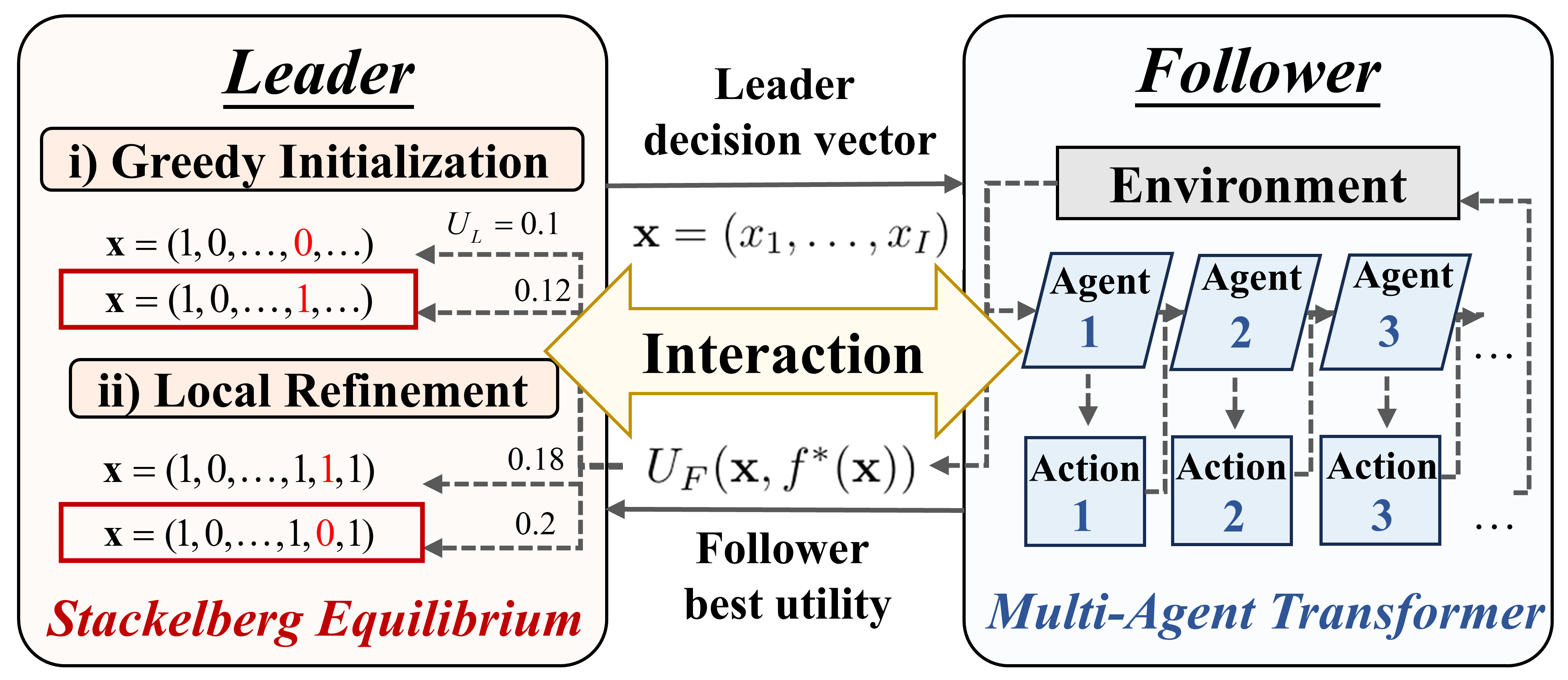}
	\caption{Leader–follower architecture of the proposed SG-MAT converging toward SE.}
	\label{fig:SG-MAT}
\end{figure}

Since the re-optimization problem is NP-hard, computing the global optimum is intractable in an online setting. We therefore develop a hierarchical Stackelberg with MAT to approximate the optimal solution. As shown in Fig.~\ref{fig:SG-MAT}, the leader generates a decision vector $\mathbf{x}$ that selects a subset of services for re-optimization, while the follower employs an MAT-based cooperative policy to compute its best response $\mathbf{f}^*(\mathbf{x})$ and report the associated re-optimization utility. Through iterative leader–follower interactions, the joint strategy pair $(\mathbf{x},\mathbf{f})$ is driven toward an approximate SE for the current re-optimization event, \del{enabling adaptive and stable re-optimization under nonlinear interference coupling} \add{enabling coordinated and consistent re-optimization under nonlinear interference coupling}.

\begin{figure}[tp]
	\centering
        \includegraphics[width=0.48\textwidth]{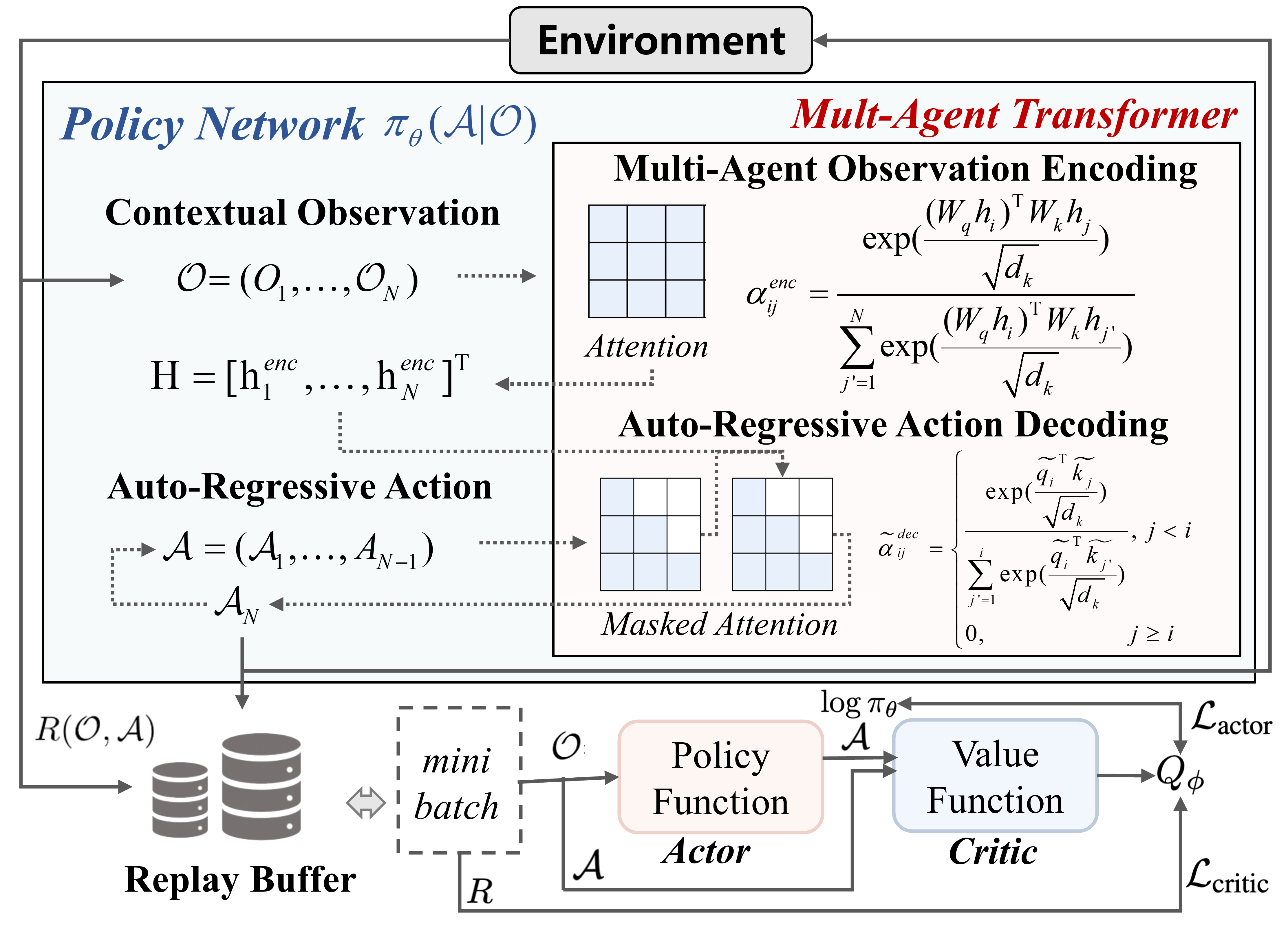}
	\caption{MAT policy framework combining attention-based coordination with training procedure.}
	\label{fig:MAT}
\end{figure}

This section first presents multi-service joint reconfigurations via MAT under nonlinear coupling \cite{J. Liu}. As shown in Fig. \ref{fig:MAT}, the MAT architecture employs an attention-based encoder–decoder to capture implicit inter-service dependencies and nonlinear interference. In addition, an efficient training procedure is developed to enable scalable and context-aware channel reassignment.

\subsection{MAT-Based Coordinated Policy Architecture}
Effective coordinated policy generation calls for a formalism that integrates agent-specific contexts with their joint effects. We therefore model the follower-layer reconfiguration as a multi-agent decision process, in which agents select actions based on contextual observations and receive a global reward. For a set of $M$ agents indexed by $\mathcal{I} = \{1, 2, \dots, M\}$, this process is characterized by the four-tuple
\begin{align}
	\mathcal{M} = \langle \mathcal{O}, \mathcal{A}, R, 
	\pi_\theta(\mathcal{A} | \mathcal{O}) \rangle,
\end{align}
\noindent where the components are defined as follows:
\begin{itemize}
	
	\item $\mathcal{O}=(\mathcal{O}_1,\dots,\mathcal{O}_M)$ denotes the global context, where each $\mathcal{O}_i=[\mathcal{L}_i,\mathcal{S}_i]\in\mathbb{R}^{2W+3}$ concatenates lightpath-state features $\mathcal{L}_i$ and service features $\mathcal{S}_i$. Specifically, $\mathcal{L}_i$ comprises two $W$-dimensional ISRS summaries over the route $\mathcal{P}_{s_i}$, namely the elementwise maximum $\max_{e\in\mathcal{P}_{s_i}} I_e(k)$ and the mean $\frac{1}{|\mathcal{P}_{s_i}|}\sum_{e\in\mathcal{P}_{s_i}} I_e(k)$ for $k=0,1, \dots, W-1$, with $I_{e}(k)=\frac{1}{|\Omega_{e}|}\sum_{j\in\Omega_{e}} r(|f_k-f_j|)$ and $\Omega_e$ representing the set of occupied channels on link $e$. In addition, $\mathcal{L}_i$ records the normalized path-length maxima and means over $\mathcal{P}_{s_i}$, while $\mathcal{S}_i$ contains the normalized service bitrate $\bar b_i$.
	
	\item $\mathcal{A} = (\mathcal{A}_1, \dots, \mathcal{A}_{M}) \in \mathcal{W}^M$ is the joint action, where $\mathcal{A}_i$ denotes agent $i$'s selected channel index;

	\item $R(\mathcal{O}, \mathcal{A})$ denotes the global reward measuring the overall effectiveness of the joint action $\mathcal{A}$ under the network context $\mathcal{O}$. It is computed as a weighted sum of utilization and GSNR improvements, as defined in~\eqref{eq:U_F}.

	\item $\pi_\theta(\mathcal{A}|\mathcal{O})$ is the joint policy, parameterized by $\theta$, outputting the probability distribution over actions given context $\mathcal{O}$.
	
\end{itemize}

The optimization objective is to learn a policy $\pi_\theta$ that, given the system context $\mathcal{O}$, selects actions $\mathcal{A} = (\mathcal{A}_1, \dots, \mathcal{A}_M)$ to maximize the expected reward:
\begin{align}
	\pi^* = \arg\max_{\pi_\theta} \ \mathbb{E}_{\mathcal{A} \sim \pi_\theta(\cdot|\mathcal{O})} \left[ R(\mathcal{O}, \mathcal{A}) \right].
\end{align}

Optimizing the joint policy $\pi_\theta(\mathcal{A}|\mathcal{O})$ is hindered by combinatorial action explosion and inter-agent coupling. To address this, we leverage the Multi-Agent Advantage Decomposition Theorem \cite{J. G. Kuba} and employ a Transformer-based architecture to explicitly model the joint policy of multiple agents \cite{M. Wen}. According to the theorem, the joint advantage function $\mathbb{A}^\pi(\mathcal{O}, \mathcal{A})$ can be exactly decomposed as:
\begin{align}
	\mathbb{A}^\pi(\mathcal{O}, \mathcal{A}) = \sum_{i=1}^{N} \mathbb{A}^\pi_i(\mathcal{O}, \mathcal{A}_{<i}, \mathcal{A}_i),
\end{align}
\noindent where $\mathbb{A}^\pi_i(\mathcal{O}, \mathcal{A}_{<i}, \mathcal{A}_i)$ denotes the marginal advantage of agent $i$ under observation $\mathcal{O}$, conditioned on $\mathcal{A}_{<i} = (\mathcal{A}_1,\dots,\mathcal{A}_{i-1})$ and the current action $\mathcal{A}_i$. This decomposition implies that, under an autoregressive decision scheme, each agent’s policy can be optimized independently based on prior decisions and the shared global context. Consequently, the joint optimization complexity can be reduced from $O(W^M)$ to $O(MW)$, significantly improving the scalability and convergence of policy learning. 

Moreover, effective decomposition relies on the ability of agents to condition their decisions on both the global context and preceding actions—an assumption that traditional frameworks struggle to model. We thus employ a Transformer encoder-decoder architecture, explicitly modeling inter-agent dependencies and ensuring structural alignment with the theorem's sequential decision-making requirement. The encoder embeds each agent's local context $\mathcal{O}_i$ via a multilayer perceptron: $\mathbf{h}_i = f_{\text{emb}}(\mathcal{O}_i)$, and employs standard self-attention, where each agent’s contextual embedding serves simultaneously as the source of queries, keys, and values, enabling the encoder to capture pairwise inter-agent relations in a permutation-equivariant manner:
\begin{align}
	\alpha_{ij}^{\text{enc}} = \frac{\exp\left( (\mathbf{W}_q\mathbf{h}_i)^\top \mathbf{W}_k\mathbf{h}_j / \sqrt{d_k} \right)}{\sum\limits_{j'=1}^{N} \exp\left( (\mathbf{W}_q\mathbf{h}_i)^\top \mathbf{W}_k\mathbf{h}_{j'} / \sqrt{d_k} \right)},
\end{align}
\begin{align}
	\tilde{\mathbf{h}}_i = \sum_{j=1}^{N} \alpha_{ij}^{\text{enc}} \mathbf{W}_v\mathbf{h}_j,
\end{align}

\noindent where $\mathbf{W}_q$, $\mathbf{W}_k$, and $\mathbf{W}_v$ are the parameter matrices. After residual connection and layer normalization, all agent representations $\mathbf{h}_i^{\text{enc}} = \text{LayerNorm}\left( \mathbf{h}_i + \text{MLP}(\tilde{\mathbf{h}}_i) \right)$ are stacked to yield the global context matrix: $\mathbf{H} = [\mathbf{h}_1^{\text{enc}}, \mathbf{h}_2^{\text{enc}}, \dots, \mathbf{h}_M^{\text{enc}}]^\top$.

The decoder then autoregressively generates actions $\mathcal{A}_i$ conditioned on both the global context matrix $\mathbf{H}$ and the previously selected actions. We define an action embedding matrix $\mathbf{Z}=[\mathbf{z}_1, \dots, \mathbf{z}_{M+1}]$, initialized with a start-of-sequence token followed by the one-hot embeddings of the already generated actions. At decoding step $i$, the partial action sequence $\mathbf{Z}_{<i}$ is first projected into query, key, and value vectors and processed through a masked self-attention block. This module aggregates information from all previously generated actions while enforcing the autoregressive constraint:
\begin{align}
	\alpha_{ij}^{\text{dec},1} = 
	\begin{cases}
		\frac{\exp\left((\mathbf{W}_q^{(1)} \mathbf{z}_i)^\top \mathbf{W}_k^{(1)} \mathbf{z}_j / \sqrt{d_k} \right)}{\sum\limits_{j'=1}^{i-1} \exp\left((\mathbf{W}_q^{(1)} \mathbf{z}_i)^\top \mathbf{W}_k^{(1)} \mathbf{z}_{j'} / \sqrt{d_k} \right)}, & j < i \\
		0, & j \geq i
	\end{cases}
\end{align}
\begin{align}
	\tilde{\mathbf{z}}_i = \sum_{j=1}^{i-1} \alpha_{ij}^{\text{dec},1} \mathbf{W}_v^{(1)} \mathbf{z}_j,
\end{align}
where $\mathbf{W}_q^{(1)}, \mathbf{W}_k^{(1)}, \mathbf{W}_v^{(1)}$ denote the projection matrices of the first decoder attention block.

To integrate global multi-agent contextual information into the decoding process, MAT further applies a second decoder attention block whose query vectors originate from the encoder outputs, while the keys and values originate from the masked self-attention state $\tilde{\mathbf{z}}_j$. This encoder-query masked attention injects inter-agent dependencies into each decoding step while maintaining the autoregressive masking over the decoder side:
\begin{align}
	\alpha_{ij}^{\text{dec},2} =
	\begin{cases}
	\frac{
		\exp \left(
		( \mathbf{W}_q^{(2)} \mathbf{h}_i^{\mathrm{enc}} )^\top
		( \mathbf{W}_k^{(2)} \tilde{\mathbf{z}}_j )
		/ \sqrt{d_k}
		\right)
	}{
		\sum\limits_{j'=1}^{i-1}
		\exp \left(
		( \mathbf{W}_q^{(2)} \mathbf{h}_i^{\mathrm{enc}} )^\top
		( \mathbf{W}_k^{(2)} \tilde{\mathbf{z}}_{j'} )
		/ \sqrt{d_k}
		\right)
	}, & j < i \\
	0, & j \geq i
	\end{cases}
\end{align}
\begin{align}
	\hat{\mathbf{z}}_i
	= \sum_{j=1}^{i-1}
	\alpha_{ij}^{\text{dec},2} \,
	( \mathbf{W}_v^{(2)} \tilde{\mathbf{z}}_j ).
\end{align}
Similarly, $\mathbf{W}_q^{(2)}, \mathbf{W}_k^{(2)}, \mathbf{W}_v^{(2)}$ are used in the second decoder attention block.

Residual connections and layer normalization refine the decoder state, after which a masked softmax output layer produces the probability distribution over agent $i$'s channel selection. The decoding procedure iterates from $i=1$ to $M$, yielding the final joint action vector 
$\mathcal{A} = ( \mathcal{A}_1, \dots, \mathcal{A}_M )$. This two-stage masked attention design enables MAT to jointly model autoregressive action dependencies and global inter-agent context, substantially improving coordination and scalability in multi-agent policy learning.

\subsection{Policy Inference and Optimization Procedure for MAT}

MAT’s effectiveness relies on accurate policy inference and optimization under dynamic conditions. As shown in Algorithm~\ref{alg:MAT}, when a blocking event is triggered, a candidate set $\mathcal{I}$ of $M$ services is selected based on bandwidth utilization. The encoder extracts inter-service dependencies from the global context $\mathcal{O}$, and the decoder autoregressively produces a coordinated joint action sequence $\mathcal{A}$. Executing these actions yields the global reward $R(\mathcal{O},\mathcal{A})$ as defined in (\ref{eq:U_F}). Each transition tuple $(\mathcal{O},\mathcal{A},R)$ is then added to the replay buffer $\mathcal{B}$, from which mini-batches of size $K$ are sampled for parameter updates.
\begin{algorithm}[htbp]
	\caption{Policy Inference and Optimization with MAT for the Re-optimization Problem}
	\label{alg:MAT}
	\begin{algorithmic}[1]
		\REQUIRE Current network topology $\mathcal{T}$, active service set $\mathcal{D}$, blocked service $s$
		\ENSURE Updated policy parameters $\theta^\star, \phi^\star$
		
		\STATE Initialization: Policy network $\pi_\theta$, critic network  $Q_\phi$, replay buffer $\mathcal{B} \gets \varnothing$
		
		\FOR{each blocking event}
		\STATE $M$ candidate services $\mathcal{I} \gets \operatorname{SelRank}(\mathcal{D}, s)$
		
		\STATE $\mathbf{H} = \operatorname{Encoder}(\mathcal{O})$
		\STATE $\mathcal{A} = \operatorname{AutoregressiveDecoder}(\mathbf{H})$
		\STATE Apply feasible actions $\mathcal{T} \gets \mathcal{A'}$ and obtain global reward $R(\mathcal{O}, \mathcal{A})$
		\STATE $\mathcal{B} \gets (\mathcal{O}, \mathcal{A}, R)$

		\IF{update interval reached}
		\STATE Sample $\mathcal{B}_{\text{mini}} \subseteq \mathcal{B}$
		\STATE Compute $\mathcal{L}_{\text{critic}}$ and $\mathcal{L}_{\text{actor}}$ according to (\ref{critic_loss}) and (\ref{actor_loss})
		\STATE Update parameters: $\theta \leftarrow \theta - \alpha_\pi \nabla_\theta \mathcal{L}_{\text{actor}}, \quad
		\phi \leftarrow \phi - \alpha_Q \nabla_\phi \mathcal{L}_{\text{critic}}$
		\ENDIF
		\ENDFOR
		
		\STATE \textbf{return} $\theta^\star, \phi^\star$
	\end{algorithmic}
\end{algorithm}

MAT employs a single-step entropy-regularized SAC-style actor–critic framework \cite{X. Gao}, optimized under the following objective:
\begin{align}
	J(\pi) =
	 \mathbb{E}_{(\mathcal{O}, \mathcal{A}) \sim \rho_\pi}
	\left[
	Q(\mathcal{O},\mathcal{A})
	- \alpha \log \pi(\mathcal{A}\mid \mathcal{O})
	\right],
\end{align}

\noindent where $\alpha$ controls the exploration-exploitation trade-off, and $Q(\mathcal{O},\mathcal{A})$ evaluates the expected single-step reward under the given context.

The twin critics $Q_{\phi_1}$ and $Q_{\phi_2}$ are trained to regress toward the instantaneous global reward. Given a mini-batch $(\mathcal{O}^{(j)},\mathcal{A}^{(j)},R^{(j)})$, the critic loss is
\begin{align}
	\mathcal{L}_{\text{critic}}
	=
	\frac{1}{K}\sum_{j=1}^{K}
	&\Big[
	 \ell(Q_{\phi_1}(\mathcal{O}^{(j)},\mathcal{A}^{(j)}), R^{(j)}) \notag \\
	& +
	\ell(Q_{\phi_2}(\mathcal{O}^{(j)},\mathcal{A}^{(j)}), R^{(j)})
	\Big],
	\label{critic_loss}
\end{align}
where $\ell(\cdot)$ denotes the smooth-$L_1$ (Huber) loss.

The actor is updated by sampling a new autoregressive joint action $\tilde{\mathcal{A}}^{(j)}$ from $\pi_\theta$ and maximizing its entropy-regularized value. The resulting actor loss is
\begin{align}
	\mathcal{L}_{\text{actor}}
	=
	\frac{1}{K}
	\sum_{j=1}^K
	\Big[
	\alpha \cdot \bar{\ell}^{(j)}
	-
	\min(Q_{\phi_1},Q_{\phi_2})
	(\mathcal{O}^{(j)},\tilde{\mathcal{A}}^{(j)})
	\Big],
	\label{actor_loss}
\end{align}
where the average log-probability $\bar{\ell}^{(j)}$ is computed as
\begin{align}
	\bar{\ell}^{(j)}
	=
	\frac{1}{M}
	\sum_{i=1}^M
	\log \pi_\theta\!\left(
	\tilde{\mathcal{A}}_i^{(j)}
	\mid
	\mathcal{O}^{(j)}, \tilde{\mathcal{A}}_{<i}^{(j)}
	\right).
\end{align}

In practice, the number of services selected for re-optimization varies across blocking events. To support mini-batch updates with consistent tensor dimensions, zero-padding and an agent-wise binary mask are applied during implementation to disable padded agents in the loss calculation.

\section{Leader Decision for Adaptive \del{and Stable} Re-optimization: Toward Stackelberg Equilibrium}

The preceding section addressed the follower-layer channel reassignment by constructing a practical approximation to the best-response policy. Building upon this, the leader layer focuses on identifying the service subset for re-optimization to maximize overall network utility. In this section, we first establish the existence of SE and then develop an efficient backward-induction-based approximation framework.

\subsection{Existence of Stackelberg Equilibrium}
\textit{Assumption A1:}
The leader’s action space $\mathcal{X} = \{0,1\}^I$ is finite.
For any given leader decision vector $\mathbf{x} \in \mathcal{X}$, the follower’s feasible strategy set is defined as

\begin{align}
	\mathbb{F}(\mathbf{x})
	= \Bigl\{\,
	& \mathbf{f}\in \mathcal{F}^I \;\Big|\;
	 \sum_{w\in\mathcal{W}} y_{s_i,w}=1,
	&& \forall\, s_i\in S_p, \notag\\
	& \sum_{i:\, e\in \mathcal{P}_{s_i}} y_{s_i,w}+a_{e,w} \le 1,
	&& \forall\, e\in E,\ \forall\, w\in \mathcal{W}, \notag\\
	& b_i \le C_{i}(\mathbf f) ,
	&& \forall s_i\in S_p
	\Bigr\}.
\end{align}

Since both the channel index set $\mathcal{W}$ and the service set $S_p$ are finite, $\mathbb{F}(\mathbf{x})$ is also finite.

\textit{Assumption A2:}
For every $\mathbf{x}\in\mathcal{X}$, the original channel assignment $\mathbf{f}^{\mathrm{ori}}$ is feasible, i.e., $\mathbf{f}^{\mathrm{ori}}\in\mathbb{F}(\mathbf{x})$.
Thus, the follower’s feasible set is never empty.

\textit{Assumption A3 (Attainability of follower best response):}
Because $\mathbb{F}(\mathbf{x})$ is finite for any $\mathbf{x}$, the follower’s utility function $U_F(\mathbf{x},\mathbf{f})$ attains its maximum. Hence,
\begin{align}
	\mathcal{B}(\mathbf{x})=\arg\max_{\mathbf{f}\in\mathbb{F}(\mathbf{x})} U_F(\mathbf{x},\mathbf{f})\neq\emptyset.
	\label{eq:bestresponse}
\end{align}

\textit{Definition 1:}
A pair $(\mathbf{x}^\ast,\mathbf{f}^\ast)$ constitutes a SE if
\begin{align}
	\mathbf{f}^\ast\in \arg\max_{\mathbf{f}\in\mathbb{F}(\mathbf{x}^\ast)} U_F(\mathbf{x}^\ast,\mathbf{f}),
	\quad
	\mathbf{x}^\ast\in \arg\max_{\mathbf{x}\in\mathcal{X}} U_L(\mathbf{x}).
\end{align}

This definition captures the hierarchical and sequential nature of the Stackelberg game: the follower reacts optimally to the leader’s strategy, and the leader anticipates this reaction in optimizing its own objective.
Equivalently, at equilibrium $(\mathbf{x}^\ast,\mathbf{f}^\ast)$, the following inequalities hold:
\begin{align}
	U_F(\mathbf{x}^\ast,\mathbf{f}^\ast) &\geq U_F(\mathbf{x}^\ast,\mathbf{f}),
	&& \forall \mathbf{f}\in\mathbb{F}(\mathbf{x}^\ast), \label{eq:followerineq}
\end{align}
\begin{align}
	U_L(\mathbf{x}^\ast) &\geq U_L(\mathbf{x}),
	&& \forall \mathbf{x}\in\mathcal{X}. \label{eq:leaderineq}
\end{align}

\textit{Theorem 1:}
Under Assumptions A1–A3, there exists at least one SE $(\mathbf{x}^\ast,\mathbf{f}^\ast)$ such that
\begin{align}
	& \mathbf{f}^\ast\in \arg\max_{\mathbf{f}\in\mathbb{F}(\mathbf{x}^\ast)}\ U_F(\mathbf{x}^\ast,\mathbf{f}), \label{eq:SE1} \\
	& \mathbf{x}^\ast\in \arg\max_{\mathbf{x}\in\mathcal{X}}\Bigl(U_F(\mathbf{x},\mathbf{f}^\ast(\mathbf{x}))-\sum_{i=1}^{I}p_i x_i\Bigr). \label{eq:SE2}
\end{align}

\textit{Proof:}
Define the induced leader-layer objective:
\begin{align}
	V(\mathbf{x})
	&\triangleq
	\max_{\mathbf{f}\in\mathbb{F}(\mathbf{x})}U_F(\mathbf{x},\mathbf{f})
	-\sum_{i=1}^{I}p_i x_i \notag \\
	&=\max_{\mathbf{f}\in\mathcal{B}(\mathbf{x})}U_F(\mathbf{x},\mathbf{f})
	-\sum_{i=1}^{I}p_i x_i.
	\label{eq:Vx}
\end{align}

From (\ref{eq:bestresponse}), the maximizer $\mathbf{f}^\ast(\mathbf{x})$ exists for each $\mathbf{x}\in\mathcal{X}$, hence $V(\mathbf{x})$ is well-defined. By Assumption A1, the leader’s decision space $\mathcal{X}$ is finite, so there exists
\begin{align}
    \mathbf{x}^\ast\in \arg\max_{\mathbf{x}\in\mathcal{X}} V(\mathbf{x}).
    \label{eq:xstar}
\end{align}

Since $\mathcal{B}(\mathbf{x}^\ast)\neq\emptyset$, select any
\begin{align}
    \mathbf{f}^\ast\in\mathcal{B}(\mathbf{x}^\ast)=\arg\max_{\mathbf{f}\in\mathbb{F}(\mathbf{x}^\ast)} U_F(\mathbf{x}^\ast,\mathbf{f}),\label{eq:fstar}
\end{align}
which guarantees follower optimality given $\mathbf{x}^\ast$.

For any other leader decision $\mathbf{x}\in\mathcal{X}$, let $\mathbf{f}^\ast(\mathbf{x})\in\mathcal{B}(\mathbf{x})$.
Then, by (\ref{eq:Vx}) and (\ref{eq:xstar}), we have
\begin{align}
	V(\mathbf{x}^\ast)
	&=U_F(\mathbf{x}^\ast,\mathbf{f}^\ast)
	-\sum_{i=1}^{I}p_i x_i^\ast \notag \\
	&\geq U_F(\mathbf{x},\mathbf{f}^\ast(\mathbf{x}))
	-\sum_{i=1}^{I}p_i x_i.
	\label{eq:ineqV}
\end{align}

Inequality (\ref{eq:ineqV}) implies that, anticipating the follower’s best response $\mathbf{f}^\ast(\mathbf{x})$, the leader has no unilateral incentive to deviate from $\mathbf{x}^\ast$.
That is,
\begin{align}
	\mathbf{x}^\ast
	\in \arg\max_{\mathbf{x}\in\mathcal{X}}
	\Bigl(U_F(\mathbf{x},\mathbf{f}^\ast(\mathbf{x}))-\sum_{i=1}^{I}p_i x_i\Bigr),
	\label{eq:leaderopt}
\end{align}
which satisfies leader optimality under follower best-response anticipation.

Combining (\ref{eq:fstar}) and (\ref{eq:leaderopt}), both follower and leader optimality conditions hold simultaneously.
Therefore, $(\mathbf{x}^\ast,\mathbf{f}^\ast)$ constitutes a SE. \hfill $\blacksquare$

\subsection{Backward Induction and Practical Approximation of Stackelberg Equilibrium}

\textit{Lemma 1:} Backward induction (BI) yields subgame-perfect Nash equilibrium (SPNE).
In any finite extensive-form game with perfect information and strictly sequential moves, the pure strategy profile produced by BI is a SPNE.

\textit{Theorem 2:} 
Under Assumptions~A1–A3, the Stackelberg game defined in Section~\ref{sec:Stackelberg Game Formulation} admits a SE, and BI yields an outcome that coincides with such an equilibrium \cite{Shen}. Specifically, there exists $(\mathbf{x}^\ast,\mathbf{f}^\ast)$ satisfying (\ref{eq:SE1}) and (\ref{eq:SE2}).

\textit{Proof:}
Apply BI to the two-stage Stackelberg game. At the unique follower stage (given $\mathbf{x}$), BI selects
\begin{align}
	\mathbf{f}^\ast(\mathbf{x})\in \mathcal{B}(\mathbf{x})=\arg\max_{\mathbf{f}\in\mathbb{F}(\mathbf{x})} U_F(\mathbf{x},\mathbf{f}).
\end{align}

At the leader root, the induced objective $V(\mathbf{x})$ is well-defined by the follower step above. BI then chooses
\begin{align}
	\mathbf{x}^\ast \in \arg\max_{\mathbf{x}\in\mathcal{X}} V(\mathbf{x})
	= \arg\max_{\mathbf{x}\in\mathcal{X}}
	\bigl[\,U_F(\mathbf{x},\mathbf{f}^\ast(\mathbf{x}))-\sum_{i=1}^I p_i x_i\,\bigr].
\end{align}

Pick any $\mathbf{f}^\ast\in \mathcal{B}(\mathbf{x}^\ast)$. By construction, $\mathbf{f}^\ast$ is a follower best response to $\mathbf{x}^\ast$, and $\mathbf{x}^\ast$ maximizes the leader’s objective under the anticipated best response $\mathbf{f}^\ast(\cdot)$. Therefore $(\mathbf{x}^\ast,\mathbf{f}^\ast)$ satisfies the SE definition. \hfill $\blacksquare$

The Stackelberg game possesses a finite action space and perfect information, ensuring the existence of a well-defined SE. However, exact BI is computationally prohibitive in online operation: it requires solving the follower best-response problem for every leader decision $\mathbf{x}\!\in\!\mathcal{X}$, where $|\mathcal{X}|=2^{|S_p|}$. This exponential growth renders global enumeration infeasible for large-scale optical networks, precluding real-time re-optimization during blocking events.

To achieve tractable yet effective decision-making, we propose a scalable approximation framework termed the \textit{Greedy-and-Local-Search Stackelberg Equilibrium} (GLSE), as shown in Algorithm~\ref{alg:glse}. GLSE preserves the two-layer decision hierarchy of the Stackelberg game by treating the MAT-based follower policy trained in Section~IV as a response, while the leader’s service selection is approximated via a structured combination of greedy construction and local search \cite{H. Huang}. GLSE proceeds in two phases:

(i) \textit{Greedy Initialization (lines 5–11).}
 We first identify all services that share at least one fiber link with the blocked request and include them in the candidate set $S_p$. These services are then examined sequentially for possible inclusion in the re-optimization set. For each candidate $s_i$, the leader decision vector is temporarily updated to $\mathbf{x}_{\mathrm{c}}$ with $\mathbf{x}_{\mathrm{c}}[i]=1$, and the follower’s MAT policy computes the corresponding reassignment and follower utility $U_F$. The insertion is accepted only if $U_L$ improves upon the current best utility by at least the acceptance threshold $\eta > 0$. This ensures that every admitted service contributes a non-negligible improvement and yields a high-quality initial solution.

(ii) \textit{Local Refinement (lines 12–22).}
The algorithm then iteratively attempts to remove individual services from the current re-optimization set. Each service currently selected for re-optimization is tentatively removed by setting its indicator to zero, and the follower utility is re-evaluated via MAT. A removal is accepted only if it increases the utility $U_L$ by at least $\eta$. The refinement iterates until no individual removal can further increase the objective by more than $\eta$. The final selection vector is therefore a 1-exchange $\eta$-local optimum with respect to the approximate Stackelberg leader objective.

\begin{algorithm}[htbp]
	\caption{Approximate Stackelberg Equilibrium via GLSE}
	\label{alg:glse}
	\begin{algorithmic}[1]
		\REQUIRE Current network topology $\mathcal{T}$, active service set $\mathcal{D}$, blocked service $s$
		\ENSURE Optimal leader selection vector $\mathbf{x}^\star$
		
		\STATE Initialization: Candidate services $S_p \gets \operatorname{SelRank}(\mathcal{D}, s)$, $\mathbf{x} = (0, 0, \ldots, 0)$, maximum reward $r \gets -\infty$
		\IF{$|S_p|=0$} \STATE \textbf{return} $\mathbf{x}$ \ENDIF

		\FOR{$i \in \operatorname{ids}(S_p)$}
		\STATE $\mathbf{x}_{\mathrm{c}} \gets \mathbf{x}; \quad \mathbf{x}_{\mathrm{c}}[i] \gets 1$
		\STATE $U_F \gets \operatorname{MAT}(\mathcal{T},\, \mathbf{x}_{\mathrm{c}},\, s)$; calculate $U_L$ by (\ref{eq:U_L})
		\IF{$U_L>r+\eta$} \STATE $\mathbf{x}\gets \mathbf{x}_{\mathrm{c}}$,\; $r \gets U_F$ \ENDIF
		\ENDFOR
		
		\STATE $\texttt{improved}\gets \texttt{true}$
		\WHILE{\texttt{improved}}
		\STATE $\texttt{improved}\gets \texttt{false}$
		\FOR{$i \in \operatorname{list}(\mathbf{x})$}
		\STATE $\mathbf{x}_{\mathrm{d}} \gets \mathbf{x}; \quad (\mathbf{x}_{\mathrm{d}})_i \gets 0$
		\STATE $U_F \gets \operatorname{MAT}(\mathcal{T},\, \mathbf{x}_{\mathrm{d}},\, s)$; calculate $U_L$ by (\ref{eq:U_L})
		\IF{$U_L>r+\eta$}
		\STATE $\mathbf{x}\gets \mathbf{x}_{\mathrm{d}}$,\; $r \gets U_F$,\; $\texttt{improved}\gets \texttt{true}$,\; \textbf{break}
		\ENDIF
		\ENDFOR
		\ENDWHILE
		\STATE $\mathbf{x}^\star \gets \mathbf{x}$
		
		\STATE \textbf{return} $\mathbf{x}^\star$
	\end{algorithmic}
\end{algorithm}

\textit{Complexity Analysis:}
From Section~\ref{sec:Stackelberg Game Formulation}, exact Stackelberg optimization requires exhaustive enumeration of
(i) leader subsets of size $2^{|S_p|}$, and
(ii) follower channel assignments of size $W^{|\mathcal{S}(\mathbf{x})|}$,
yielding a worst-case complexity $\mathcal{O}\!\left(2^{|S_p|} \cdot W^{|\mathcal{S}(\mathbf{x})|} \cdot \overline{|\mathcal{P}|}\right)$. 
The proposed GLSE significantly reduces the complexity of both layers.
On the leader side, the greedy construction examines at most $|S_p|$ candidate insertions, each requiring one follower evaluation.
In the local refinement phase, each selected service can be removed at most once, incurring at most another $|S_p|$ follower evaluations.
Hence, the total number of follower calls is bounded by $2|S_p|$, and the leader-layer computational cost is $\mathcal{O}(|S_p|)$. On the follower side, the exponential search over $W^{|\mathcal{S}(\mathbf{x})|}$ assignments is replaced by a single forward pass of the MAT policy and a QoT-based utility evaluation.
The MAT encoder--decoder involves self-attention and autoregressive masked attention over $|\mathcal{S}(\mathbf{x})|$ selected services, yielding a quadratic complexity in $|\mathcal{S}(\mathbf{x})|$, while the GSNR-based QoT evaluation scales linearly with the average path length $\overline{|\mathcal{P}|}$. Thus, the cost of a single follower inference satisfies $T_F(|\mathcal{S}(\mathbf{x})|)=\mathcal{O}\!\left(|\mathcal{S}(\mathbf{x})|^{2} + \overline{|\mathcal{P}|}\cdot |\mathcal{S}(\mathbf{x})|\right).$
Combining these results, the overall complexity of GLSE is $\mathcal{O}\bigl(|S_p|\cdot T_F(N)\bigr)$, which is polynomial in both $|S_p|$ and $N$, in sharp contrast to the exponential complexity of exact Stackelberg computation.
This reduction enables real-time re-optimization during blocking events while preserving the hierarchical leader--follower structure and equilibrium-guided decision process.

\section{Performance Evaluation}
\subsection{Experimental Settings}
To validate the effectiveness of SG-MAT, we conduct extensive simulations on two representative backbone network topologies—USNET and CORONET—whose basic characteristics are summarized in Table \ref{table:topology_info}. The network operates under a C6T + L6T configuration, where each band provides an effective bandwidth of 6~THz. In our simulations, the C-band is modeled over 184.325–190.325~THz and the L-band over 190.675–196.675~THz, each comprising 80 WDM channels with 150~GHz spacing and 100~km spans. The transmission medium is standard single-mode fiber (SMF) with frequency-dependent attenuation $\alpha_j= \alpha_2\cdot(f_j-f_{ref})^2 + \alpha_1\cdot(f_j-f_{ref}) + \alpha_0$, where $\alpha_0=0.162~dB/km$, $\alpha_1=-7.3764 \times 10^{-5}  dB/(km · nm)$, $\alpha_2=3.7685\times 10^{-6} dB/(km\cdot nm^2)$, where $f_{ref} = c/\lambda_{ref}$ denotes the reference frequency corresponding to $\lambda_{ref}=1550$~nm, and $c$ is the speed of light in vacuum. $\beta_2=-21.7ps^2/km$, $\beta_3=0.144ps^3/km$, $\gamma=1.21(W\cdot km)^{-1}$, $\varepsilon=1$, $h=6.626\times10^{-34}$, $C_r=0.028/W/km/THz$. The amplifier spontaneous emission factor $n_{sp}$ is set as $\mathrm{NF}_{\mathrm{lin}}/2$ with $\mathrm{NF}_{\mathrm{dB}}=4.5$. Launch powers are 5.9~dBm (C-band) and 4.4~dBm (L-band) \cite{T. Gerard}. 
Channel capacity is dynamically determined by GSNR, using $f_{\text{mod}}(\cdot)$ function from \cite{A. Souza}. All other parameters are set consistently with \cite{C. Lasagni,D. Semrau}.
\add{These settings follow representative multi-band optical-network studies and are consistent with practical system assumptions for C+L-band transmission.}

\begin{table}[bp]
	\centering
	\caption{Topological Metrics of USNET and CORONET}
	\label{table:topology_info}
	\begin{tabular}{lcc}
		\hline
		\textbf{Metric} & \textbf{USNET} & \textbf{CORONET} \\
		\hline
		Number of Nodes & 24 & 75 \\
		Number of Links & 43 & 99 \\
		Average Node Degree & 3.58 & 2.93 \\
		Average Shortest Path Length (km) & 2990 & 5527 \\
		\hline
	\end{tabular}
\end{table}

In the Stackelberg model, the re-optimization cost in \eqref{eq:U_L} is taken identical for all services for simplicity, i.e., \(p_i = 0.01\). The weighting parameter in \eqref{eq:U_F} is set to \(k = 0.02\), chosen based on the relative magnitudes of utilization and GSNR gain under the network conditions, and to emphasize the dominant role of utilization in the follower utility.

\del{Services are randomly generated between uniformly chosen source–destination pairs. The probability of a service with bitrate \( R_k \in \{100, 200, \dots, 600\} \)~Gbps is determined using exponential weighting, with the probability mass function given by $p_k \propto \exp(-\beta R_k)$, where \( \beta = 0.005 \). This results in a traffic pattern where low-rate inference flows are more frequent than high-rate training tasks, aligning with typical AIDC workloads.}

\add{Services are generated between uniformly selected source--destination pairs, with bitrates chosen from \(R_k \in \{100, 200, \dots, 600\}\)~Gbps. Instead of assuming a uniform bitrate mix, we adopt an exponential weighting model motivated by Alibaba's GPU pod and DLRM inference traces~\cite{alibaba_trace}. These traces indicate a skewed workload-scale distribution, where small- and medium-scale workloads occur more frequently than large-scale ones, consistent with frequent inference-serving workloads and fewer but more resource-intensive training-oriented tasks~\cite{B. Sun}. After mapping the trace-derived workload scales to the six bitrate classes, the resulting distribution exhibits a trend close to \(p_k \propto \exp(-\beta R_k)\) with \(\beta=0.005\), motivating its use for bitrate sampling.}

The MAT module consists of ten encoder layers and 24 attention heads with 576-dimensional embeddings. Both actor and critic adopt two-layer fully connected networks (256 hidden units per layer) with learning rates of $1\times10^{-4}$ and $5\times10^{-4}$, batch size $K=128$, discount factor $\delta = 0.99$, and entropy weight $\lambda_{\mathrm{ent}} = 0.02$. 
\add{
The MAT policy is trained on USNET and then evaluated on both USNET and CORONET without retraining or fine-tuning. This cross-topology evaluation is supported by MAT's topology-independent observation design. Instead of using topology identifiers, absolute node labels, or fixed adjacency matrices, MAT represents each candidate service through route-level link-state statistics and service attributes. This maps services from different topologies into a common resource- and physical-layer feature space, allowing the policy to learn transferable patterns rather than topology-specific node or link dependencies.
}
All experiments are executed on an Apple M1~Pro platform. Each main test is conducted with 100,000 requests for performance evaluation.

To ensure comprehensive evaluation, comparisons are performed at both module and system levels. The module-level experiments assess the effectiveness of the proposed MAT, while the system-level experiments evaluate the overall SG-MAT framework.
At the module level, three baseline schemes are considered:  
(i) \textit{Random Fit (RF)}—shortest-path routing with random-fit channel assignment and no re-optimization; 
(ii) \textit{SDFA}—defragmentation-driven allocation \cite{B. Bao} based on the network fragmentation metric~\cite{S. Ba}; and  
(iii) \textit{DRL}—a single-agent deep reinforcement learning model sharing the same single-agent state, action, and reward spaces as MAT.  
At the system level, SG-MAT is compared against \del{three} \add{four} online optimization baselines:
\del{
\textit{DeepDefrag} \cite{E. Etezadi}, \textit{PORL} \cite{S. Emara}, and \textit{SDORL} \cite{J. Li}, representing fragmentation-driven, Pareto-adaptive, and self-distillation–enhanced online DRL frameworks, respectively.
}
\add{
(i) \textit{DeepDefrag}~\cite{E. Etezadi}---a fragmentation-driven ORL method;
(ii) \textit{PORL}~\cite{S. Emara}---a Pareto-adaptive ORL method;
(iii) \textit{SDORL}~\cite{J. Li}---a self-distillation-enhanced ORL method; and
(iv) \textit{MPDORL}~\cite{Ayepah}---a mutual-policy-distillation ORL method with bidirectional knowledge transfer.
}

To provide an objective and multidimensional performance assessment, six quantitative metrics are employed:

\begin{itemize}	
	\item \textit{Utilization Gain} — the absolute average per-link improvement in bandwidth utilization relative to the RF baseline;
	
	\item \textit{GSNR Gain} — the absolute average per-link improvement in GSNR relative to the RF baseline;
	
	\item \textit{Blocking Ratio} — the proportion of blocked services to total requests, measuring overall accessibility;
	
	\item \textit{Re-optimization Rate} — the average number of re-optimization triggers per allocation, indicating intervention frequency;
	
	\item \textit{Re-optimization Efficiency} — ratio of successfully reconnected services to total re-optimization attempts, where a reconnection is considered successful if the blocked service is accepted after re-optimization and all affected services satisfy the QoT constraint;	
	
	\item \textit{Execution Time} — average computational time per service, reflecting real-time responsiveness and algorithmic complexity.
	
\end{itemize}

\begin{figure}[tp]
	\centering
	\subfigbottomskip=2pt
	\subfigcapskip=-5pt
	\subfigure[]{
			\includegraphics[width=0.47\linewidth]{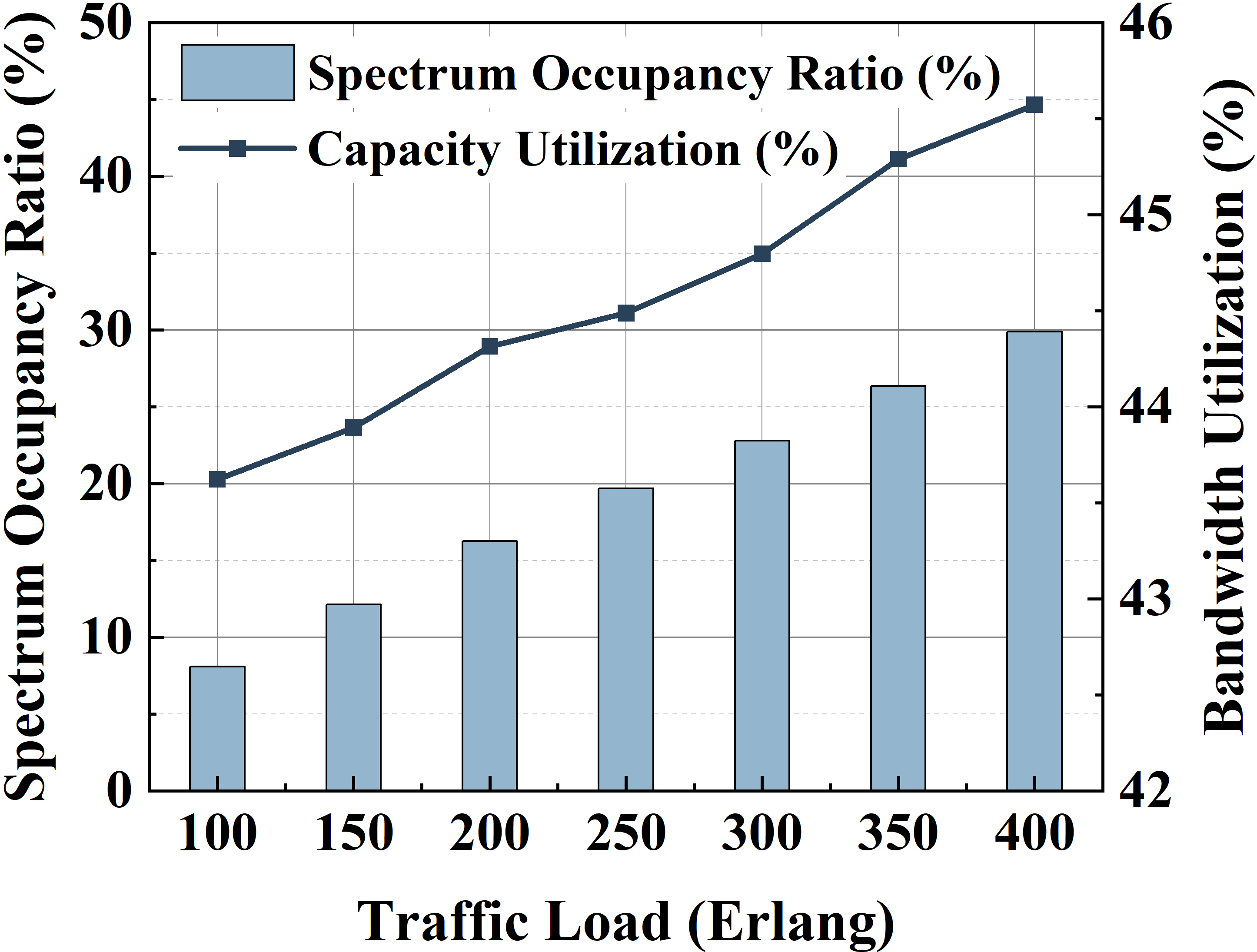}}
	\subfigure[]{
			\includegraphics[width=0.47\linewidth]{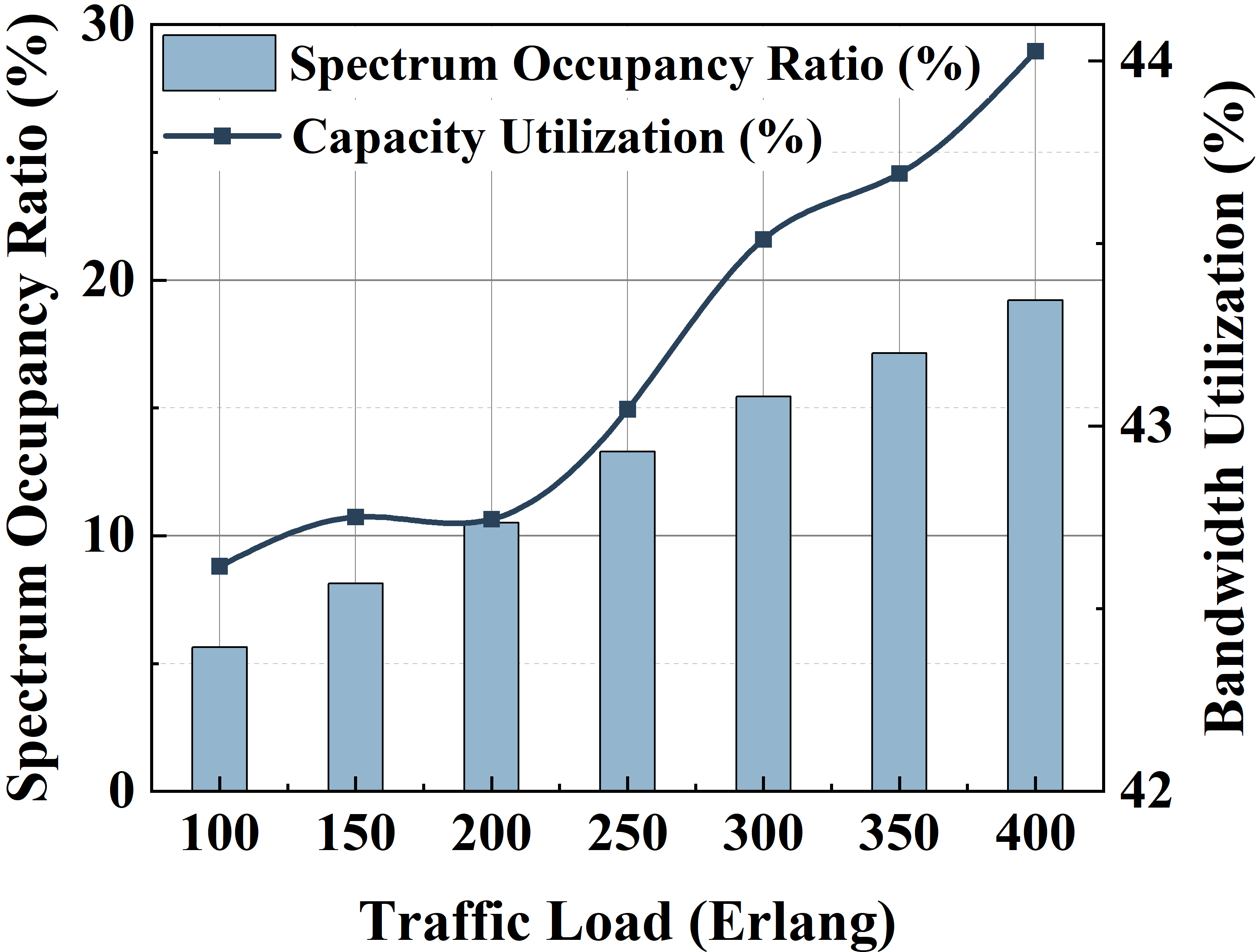}}
	\caption{The evaluations of resource occupations under topologies of (a) USNET and (b) CORONET.}
        \label{result_fig1}
\end{figure}

To determine the operational load range, Fig.~\ref{result_fig1} plots the spectrum occupation ratio and bandwidth utilization of USNET and CORONET under RF allocation. The spectrum occupation ratio denotes the proportion of occupied channels, while bandwidth utilization represents the ratio of transmitted bitrate to theoretical capacity. As traffic load increases from 100 to 400~Erlangs, both metrics rise monotonically, covering typical load regimes from light to heavy. This range is therefore adopted as the unified baseline for subsequent experiments.

\begin{figure}[tp]
	\centering
	\subfigbottomskip=2pt
	\subfigcapskip=-5pt
	\subfigure[]{
			\includegraphics[width=0.47\linewidth]{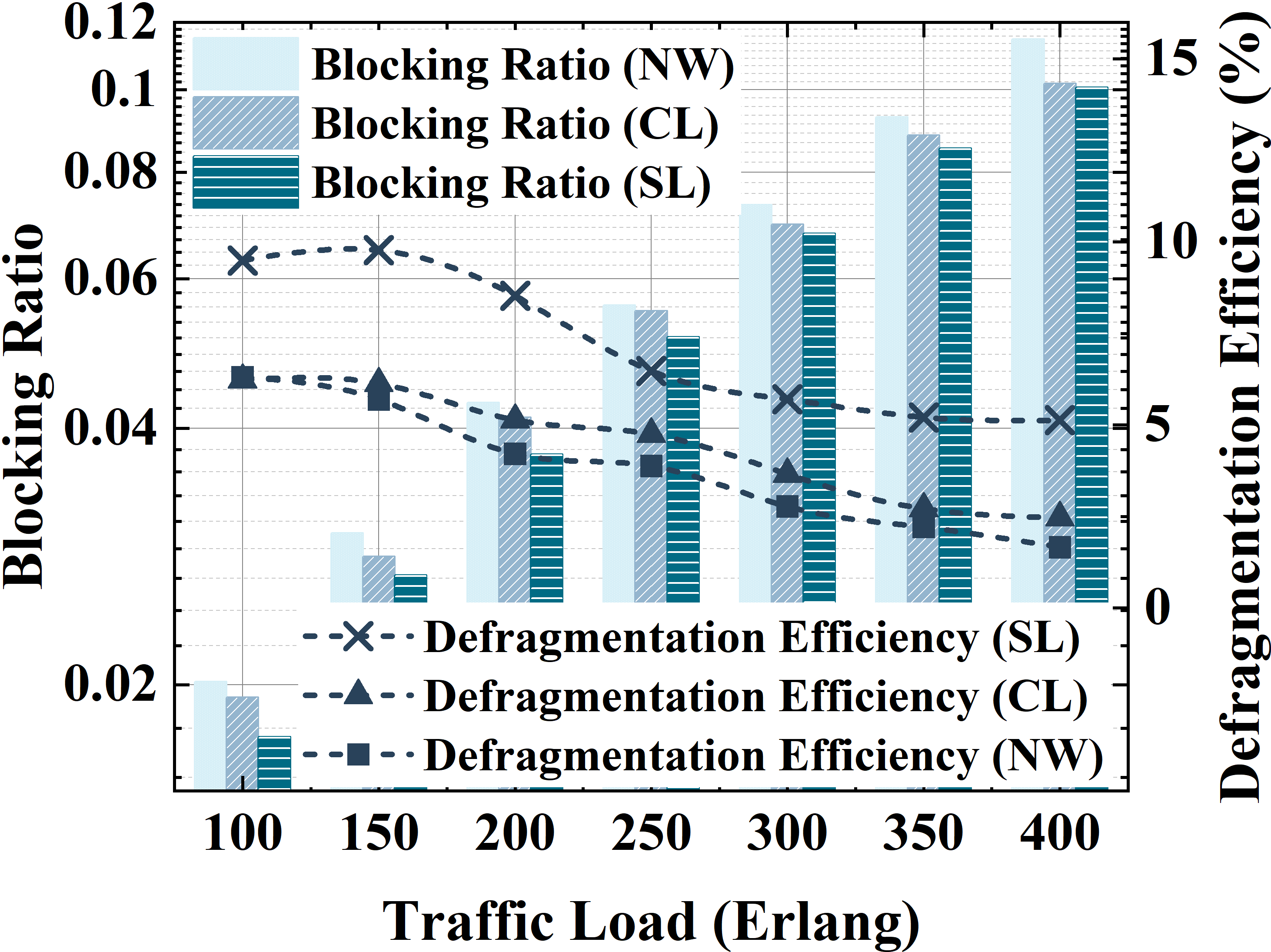}}
	\subfigure[]{
			\includegraphics[width=0.47\linewidth]{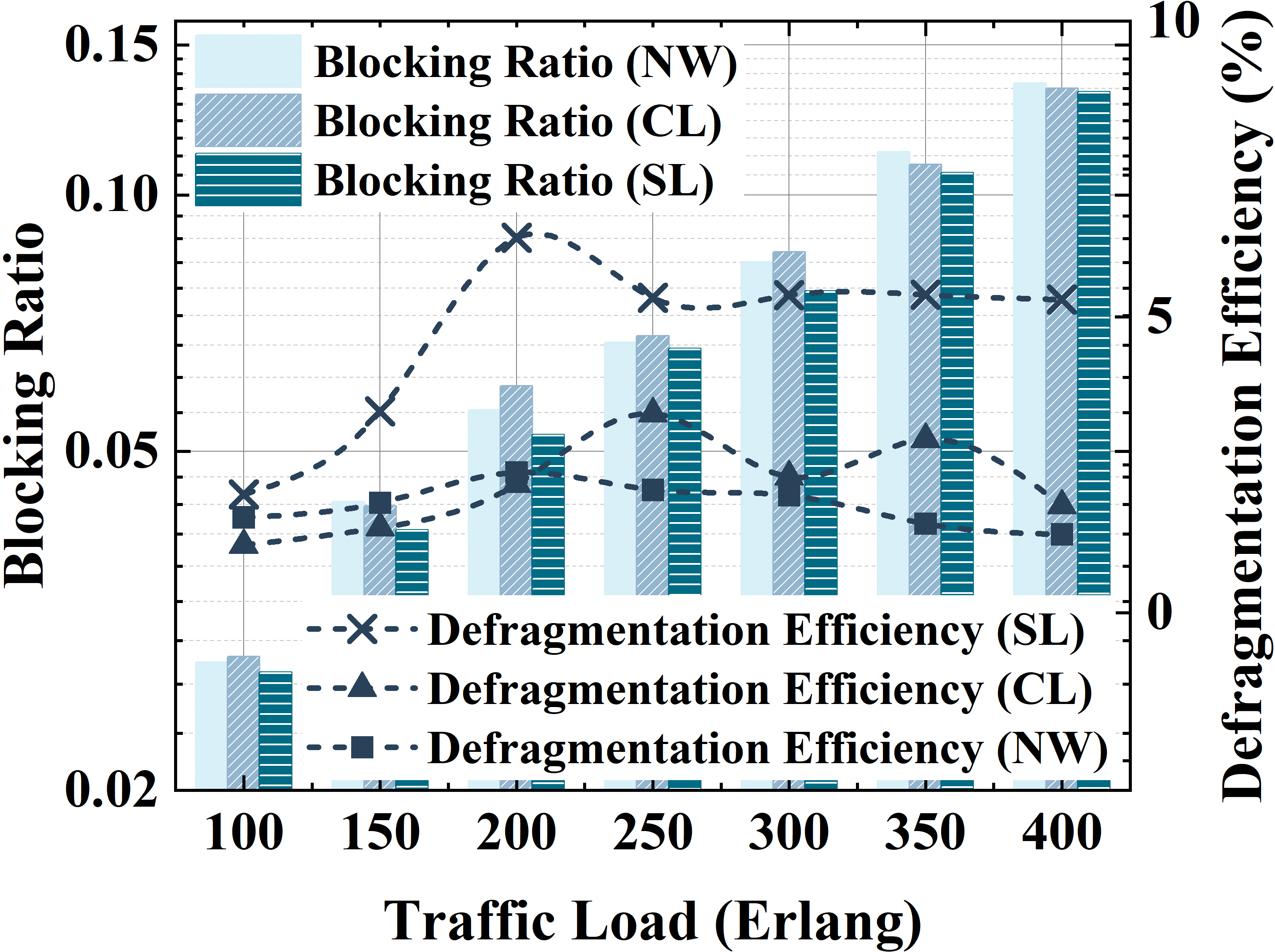}}
	\caption{Comparison of blocking ratio and re-optimization efficiency under varying service selection ranges under (a) USNET and (b) CORONET.}
        \label{result_fig2}
\end{figure}

To analyze the re-optimization scope, three candidate strategies are evaluated:  
(1) \textit{Shared Links (SL)} — services sharing at least one link with the blocked flow;  
(2) \textit{Critical Links (CL)} — services located on critical network links;  
(3) \textit{Network-Wide (NW)} — all active services.  
Upon a blocking event, up to 30 candidate services within the chosen scope are selected in ascending order of bandwidth utilization, and reallocated using a First-Fit scheme. 
As shown in Fig.~\ref{result_fig2}, SL consistently achieves the lowest blocking ratio and highest re-optimization efficiency, indicating its superior balance between performance gain and computational cost. CL and NW offer larger search spaces but suffer from stronger interference coupling \del{and unstable convergence}\add{, which reduces the efficiency of re-optimization}. Hence, the SL configuration is adopted in all subsequent evaluations.

\subsection{Performance Evaluation of the follower MAT Mechanism}

To evaluate the effectiveness of MAT, up to 30 services are selected in ascending order of bandwidth utilization upon each blocking event and reallocated using three schemes: SDFA, DRL, and the proposed MAT.

\subsubsection{Evaluations on utilization and GSNR gain}
\begin{figure}[tp]
	\centering
	\subfigbottomskip=2pt
	\subfigcapskip=-5pt
	\subfigure[]{
			\includegraphics[width=0.47\linewidth]{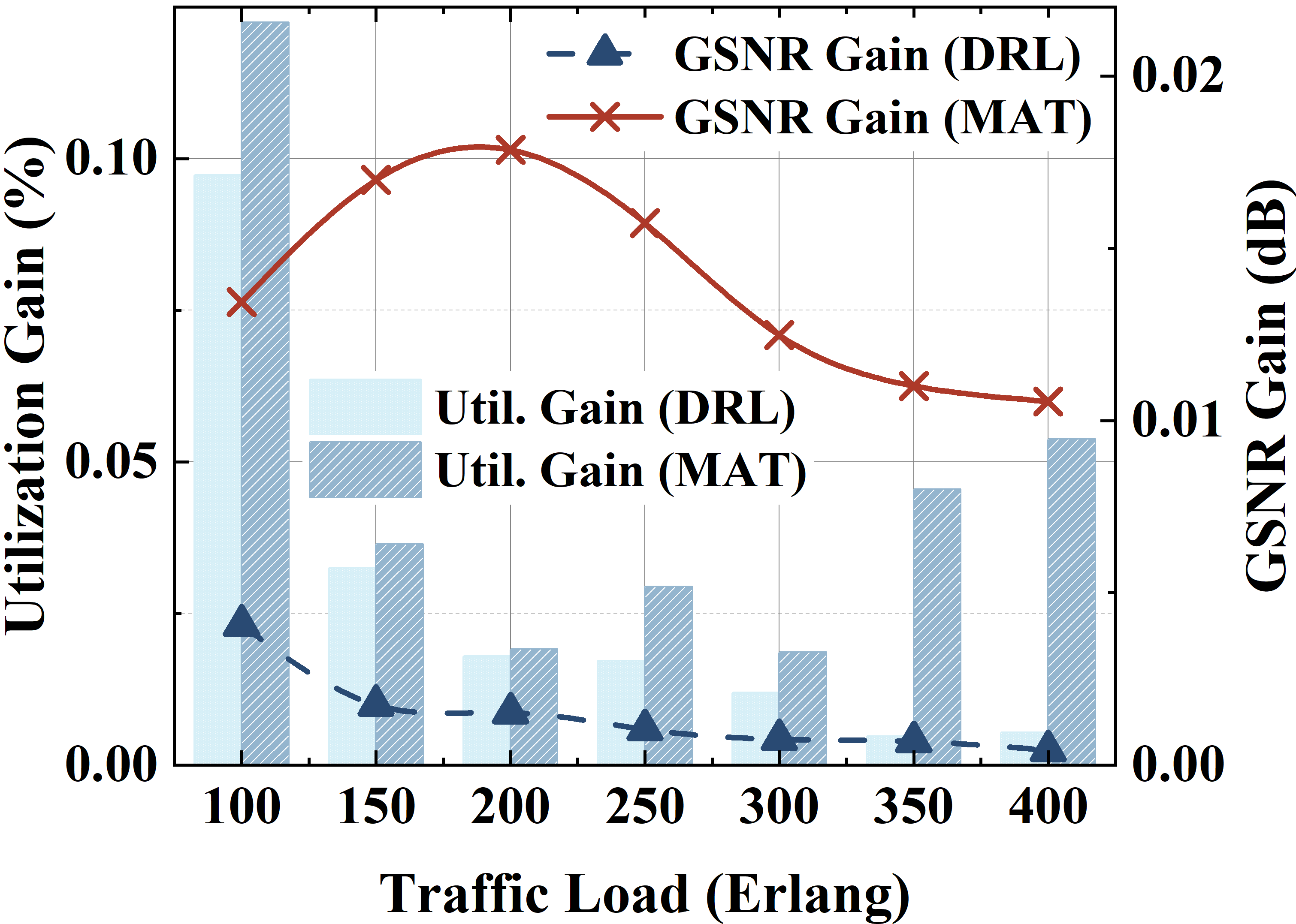}}
	\subfigure[]{
			\includegraphics[width=0.47\linewidth]{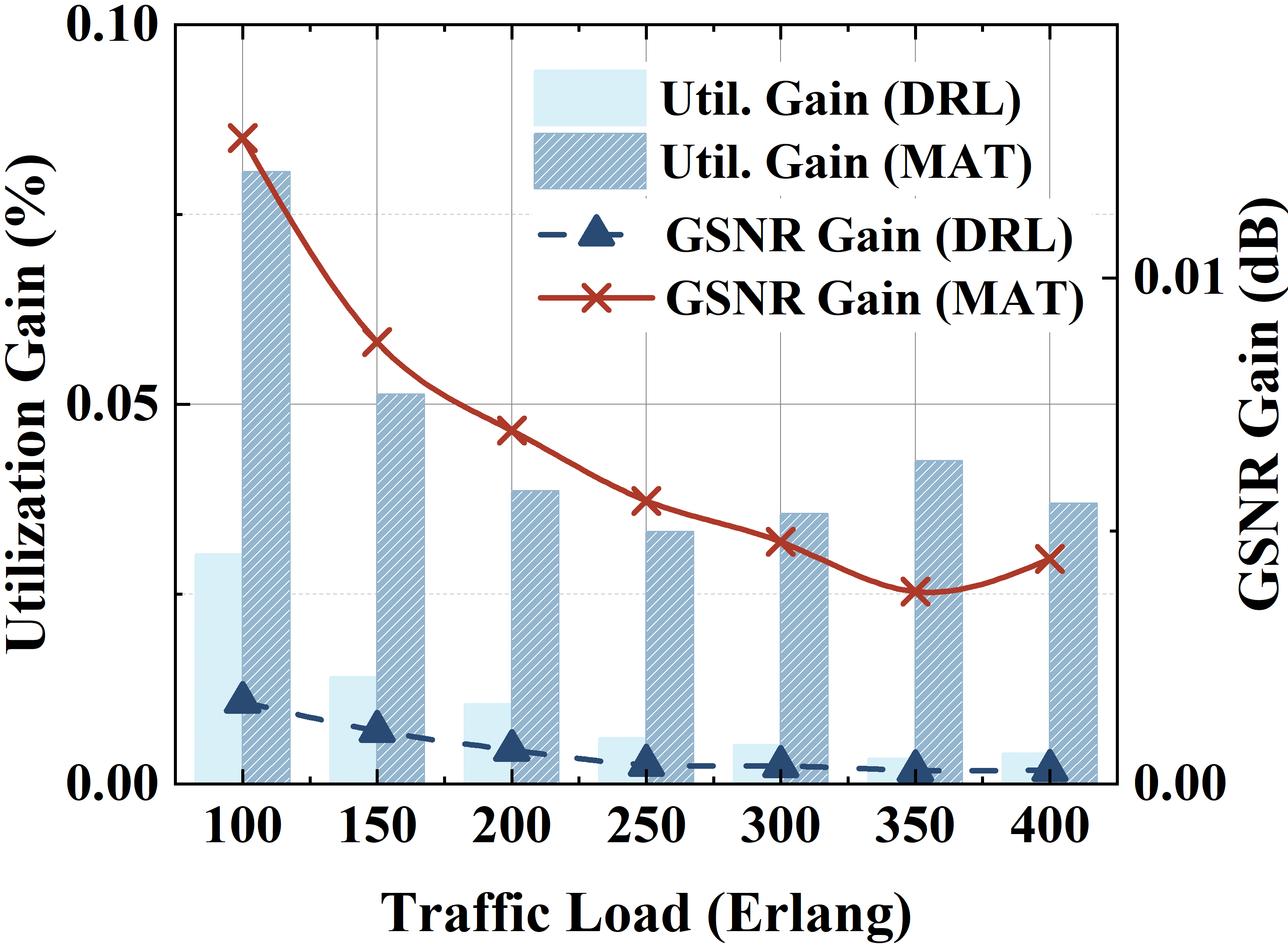}}
	\caption{Average utilization and GSNR gains of different follower schemes under (a) USNET and (b) CORONET.}
        \label{result_fig3}
\end{figure}

Fig.~\ref{result_fig3} compares the average utilization and GSNR gains achieved by DRL and MAT under both topologies. SDFA results are omitted for clarity since it primarily minimizes fragmentation rather than directly optimizing utilization or GSNR, resulting in high variability. As traffic load increases, the available spectrum shrinks and nonlinear interference intensifies, causing both metrics to gradually decline. Under light loads (100–150~Erlangs), MAT effectively exploits residual bandwidth and power margins to achieve cooperative reallocation, yielding a utilization gain of 0.05–0.10\% and a GSNR gain of 0.01–0.02~dB. In the moderate-load region (200–300~Erlangs), both metrics decrease but MAT consistently outperforms DRL. At heavy loads ($\ge$350~Erlangs), a slight rebound in utilization gain is observed, indicating that capturing and exploiting inter-service coupling enables more coordinated channel reassignments that unlock residual spectral resources, while GSNR gain remains nearly constant due to power limitations and enhanced nonlinear coupling.

\subsubsection{Evaluations on blocking ratio}

\begin{figure}[tp]
	\centering
	\subfigbottomskip=2pt
	\subfigcapskip=-5pt
	\subfigure[]{
			\includegraphics[width=0.47\linewidth]{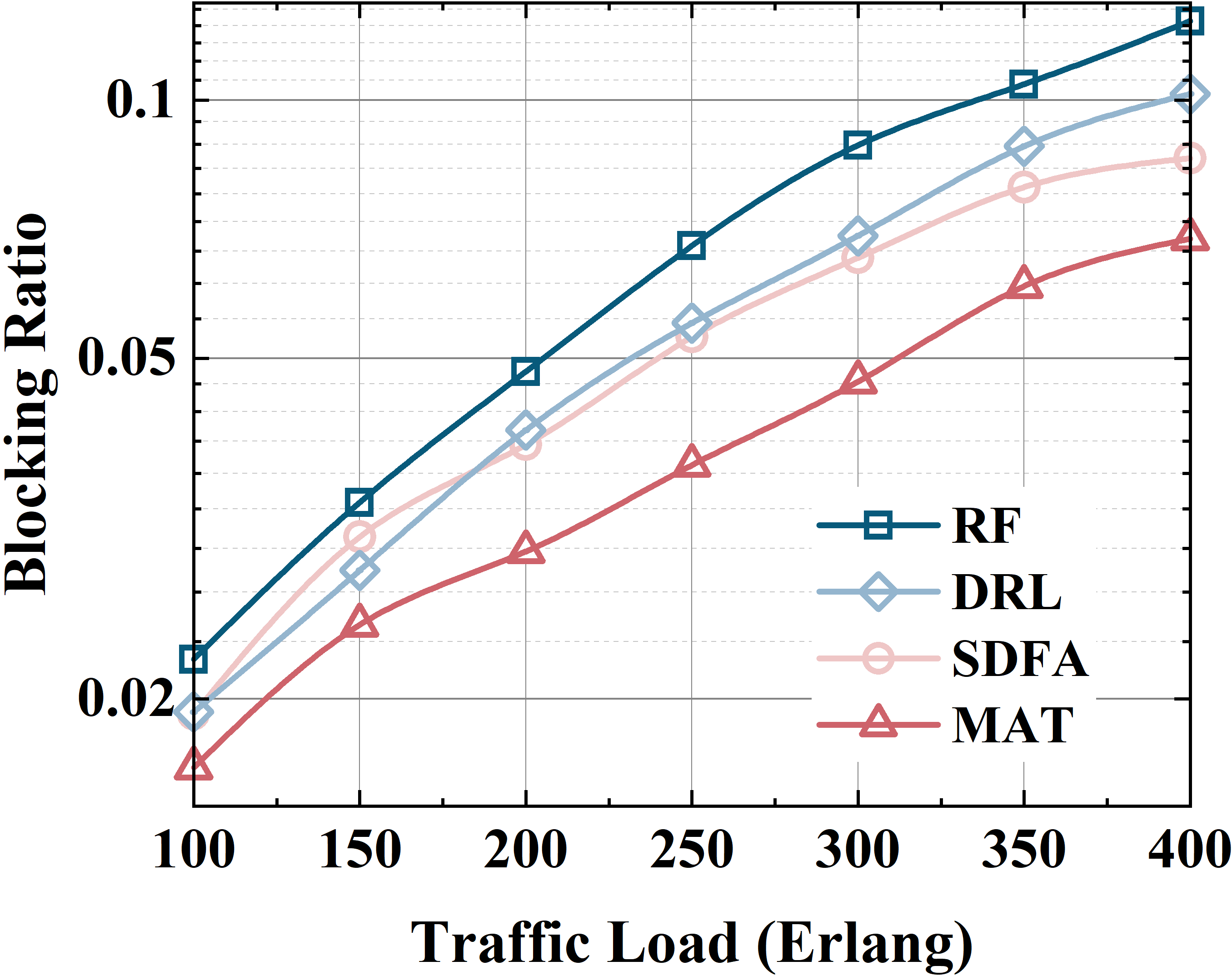}}
	\subfigure[]{
			\includegraphics[width=0.47\linewidth]{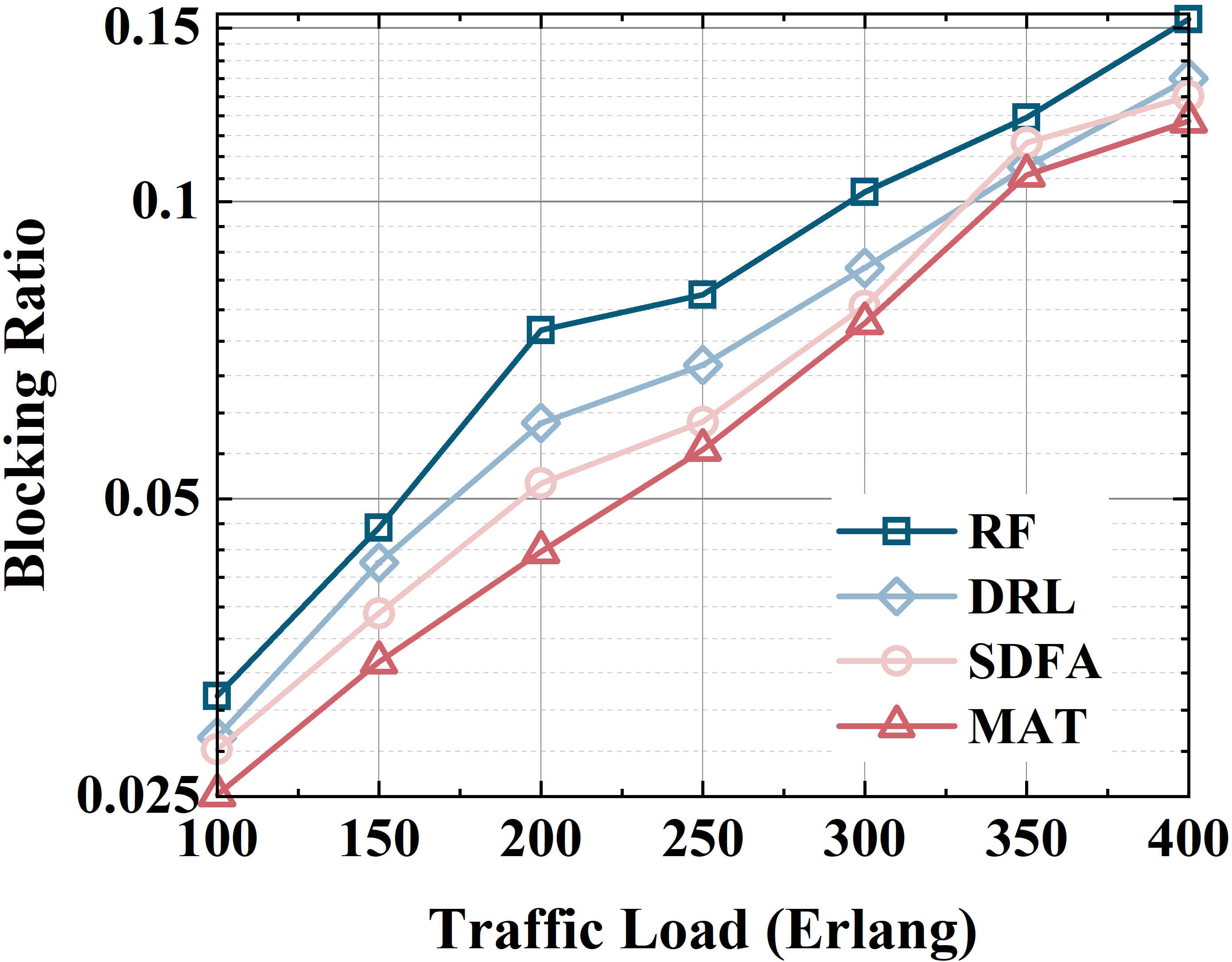}}
	\caption{Blocking ratio performance of different follower schemes under (a) USNET and (b) CORONET.}
        \label{result_fig4}
\end{figure}

Fig.~\ref{result_fig4} illustrates the blocking ratio with increasing traffic load. In both topologies, blocking increases monotonically as spectrum availability decreases and interference coupling strengthens. CORONET shows slightly higher blocking due to its longer average distance, which amplifies accumulated impairments.

In USNET, all algorithms perform comparably under light loads (100-150~Erlangs), when spectral resources are abundant and the performance gains of MAT over the baselines are relatively modest. Under moderate to heavy loads (200–400~Erlangs), MAT achieves the lowest blocking ratio, reducing it by 29.3\% and 24.5\% compared with DRL and SDFA, respectively. This confirms that MAT captures cross-link interference dependencies and enables coordinated resource re-optimization, maintaining high service accessibility.  

\add{Since the MAT policy is trained on USNET and evaluated on CORONET without retraining, the CORONET results also provide an indication of cross-topology transfer.} In CORONET, MAT still achieves lower blocking (19.5\% and 12.3\% reductions relative to DRL and SDFA) under light-to-moderate loads (100–200~Erlangs). At higher loads, the tighter coupling among long-path services constrains optimization flexibility, slightly diminishing MAT’s relative advantage. In both topologies, SDFA generally outperforms DRL due to its heuristic fragmentation-based reassignment, yet lacks explicit interference modeling and global coordination, leading to weaker performance than MAT. \add{This transferability is supported by the topology-independent observation design.}

\subsubsection{Evaluations on re-optimization costs}

\begin{figure}[tp]
	\centering
	\subfigbottomskip=2pt
	\subfigcapskip=-5pt
	\subfigure[]{
			\includegraphics[width=0.47\linewidth]{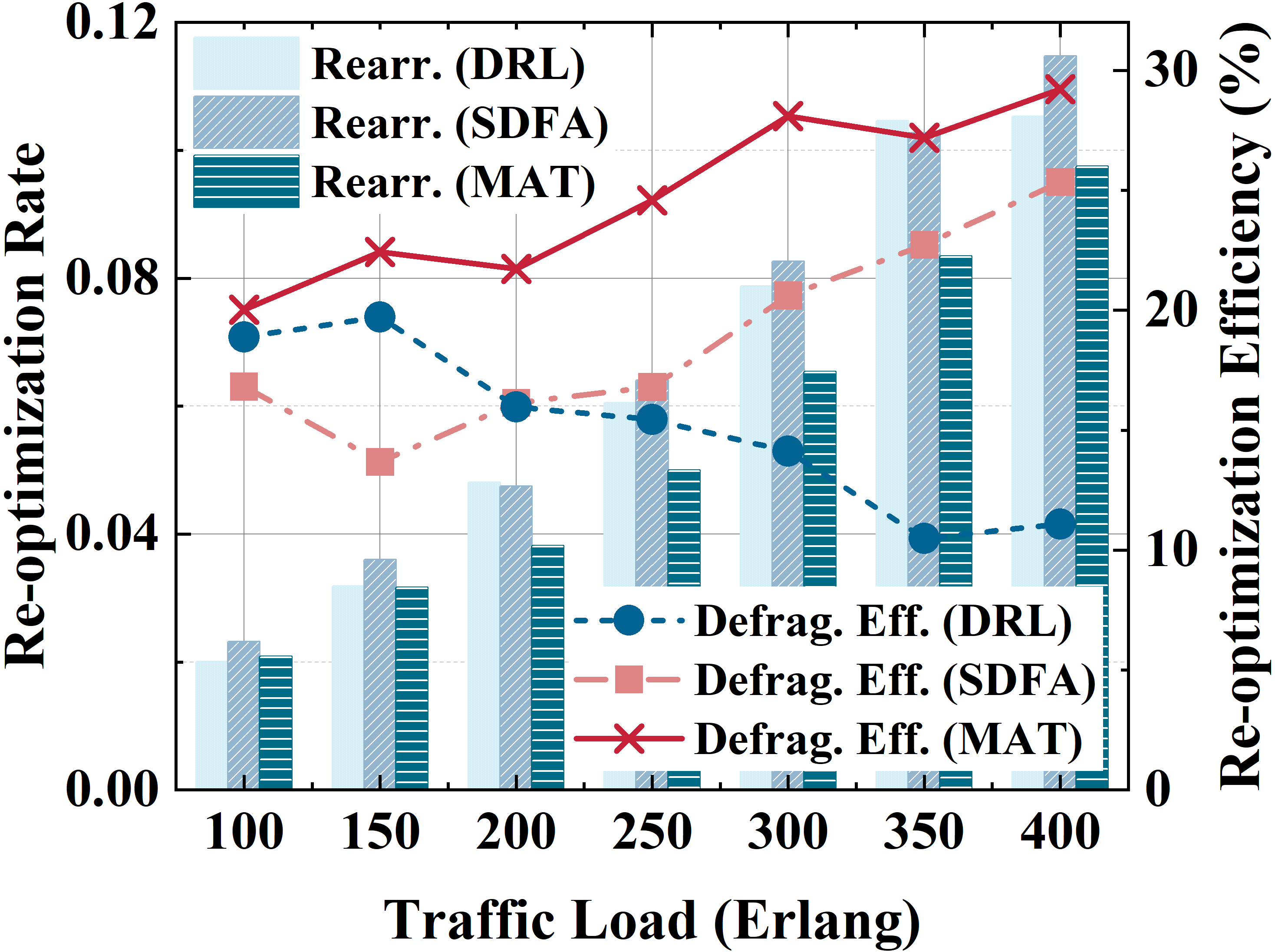}}
	\subfigure[]{
			\includegraphics[width=0.47\linewidth]{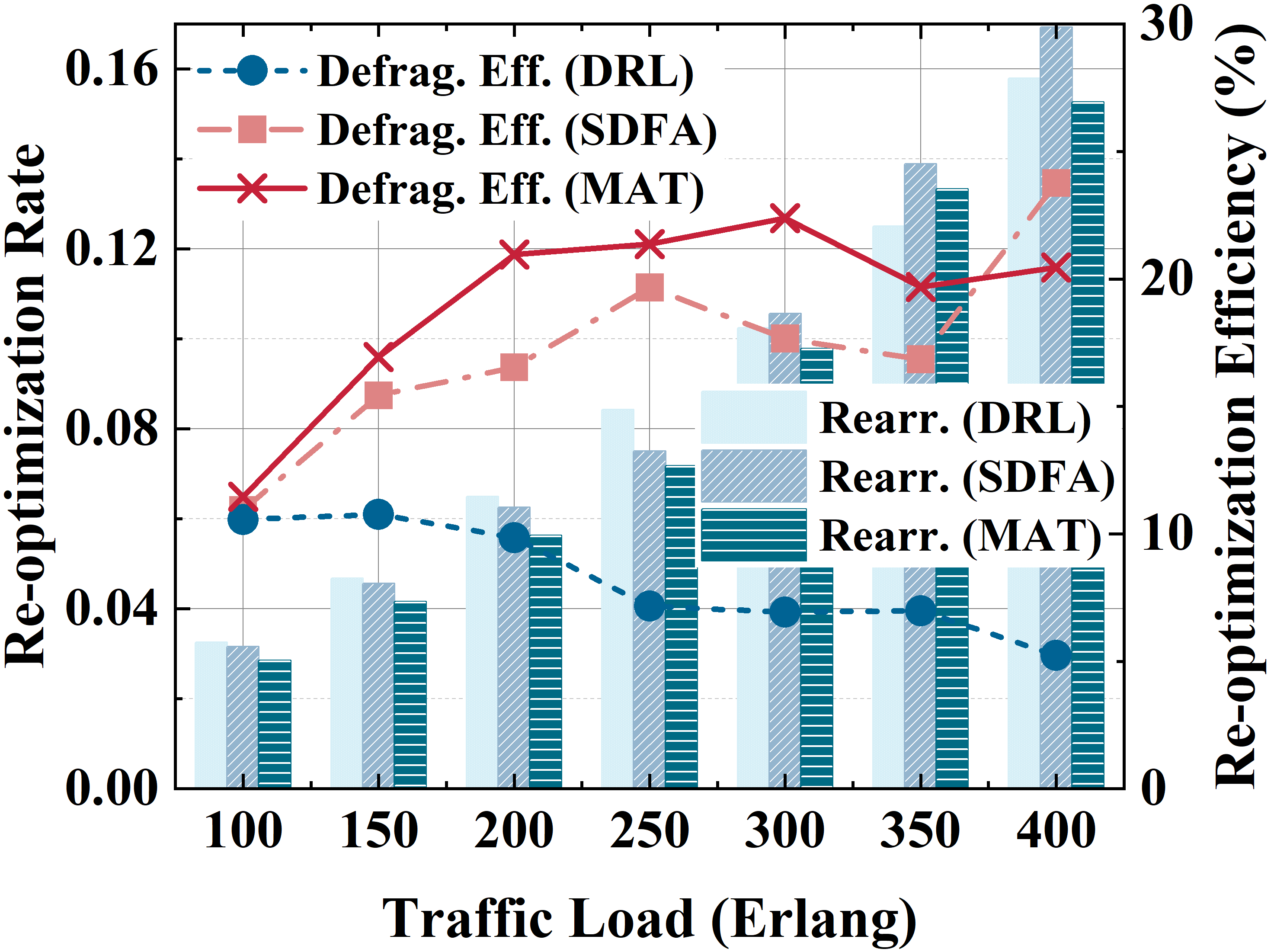}}
	\caption{Re-optimization frequency and efficiency of different follower schemes under (a) USNET and (b) CORONET.}
        \label{result_fig5}
\end{figure}

Fig.~\ref{result_fig5} shows the evolution of re-optimization rate and efficiency. With increasing traffic intensity, re-optimization events become more frequent, while efficiency diverges among schemes: MAT and SDFA show steady improvement, whereas DRL degrades progressively. Benefiting from its cooperative multi-agent structure, MAT reduces unnecessary reconfigurations \del{while maintaining stable convergence under high interference} \add{and yields more consistent re-optimization outcomes under strong interference}. DRL’s single-agent nature fails to capture inter-service dependencies, often converging to suboptimal local minima and resulting in frequent yet ineffective reconfigurations. SDFA provides local defragmentation but lacks cross-service coordination, limiting its adaptability in nonlinear interference.

\begin{figure}[tp]
	\centering
	\subfigbottomskip=2pt
	\subfigcapskip=-5pt
	\subfigure[]{
			\includegraphics[width=0.47\linewidth]{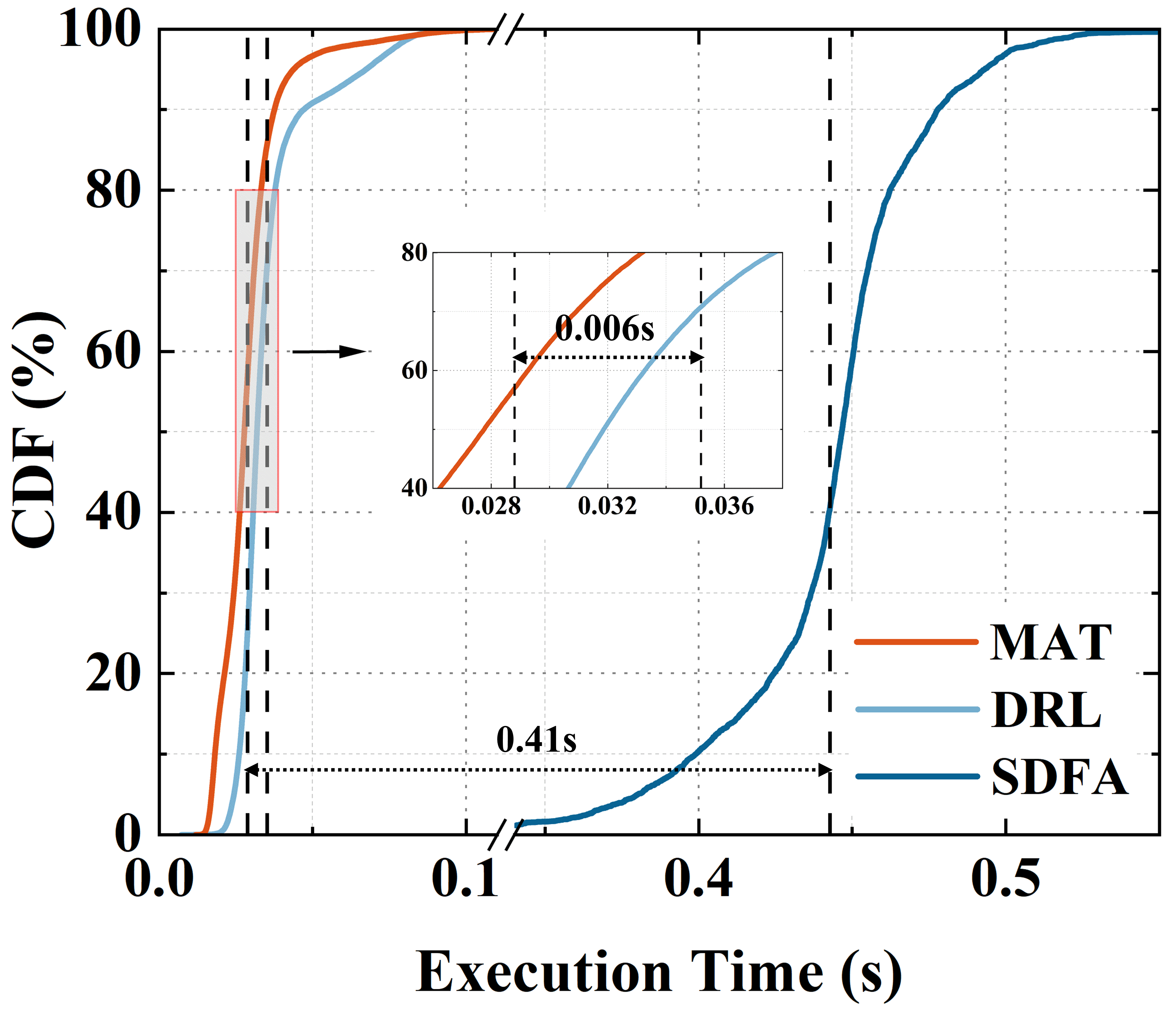}}
	\subfigure[]{
			\includegraphics[width=0.47\linewidth]{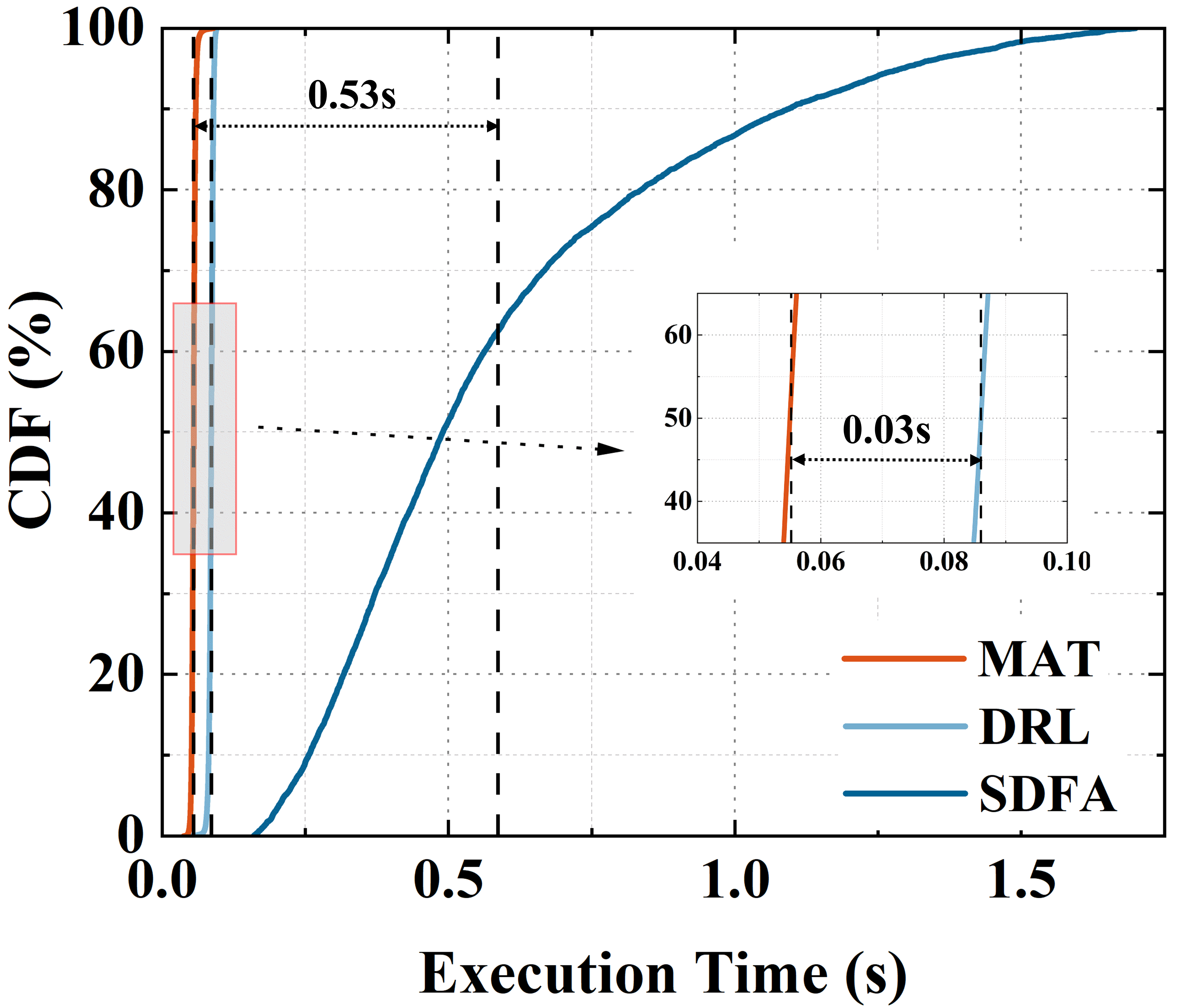}}
	\caption{CDF of the execution time of different follower schemes under (a) USNET and (b) CORONET.}
        \label{result_fig6}
\end{figure}

Fig.~\ref{result_fig6} presents the cumulative distribution function (CDF) of execution time. MAT achieves the lowest latency, while DRL incurs a moderate overhead—approximately +6~ms and +25~ms for USNET and CORONET, respectively. SDFA exhibits the highest latency due to heuristic search and exhaustive iteration.

Overall, MAT consistently delivers superior performance in bandwidth utilization, blocking suppression, and re-optimization efficiency. These results indicate that attention-based multi-agent cooperative modeling is effective in capturing implicit interference dependencies and nonlinear coupling, overcoming the limitations of independent optimization. The demonstrated scalability and consistent performance of MAT provide a robust foundation for higher-level decision-making and system-wide re-optimization within the SG-MAT framework.

\subsection{Performance Evaluation of SG-MAT}

To comprehensively assess the proposed SG-MAT framework, we also adopt a reactive setting. For each blocking event, up to 30 ongoing services with the lowest bandwidth utilization are selected as candidates, and \del{four} \add{five} algorithms—DeepDefrag, PORL, SDORL, \add{MPDORL,} and SG-MAT—are compared.

\subsubsection{Evaluations on utilization and GSNR gain}

\begin{figure}[tp]
	\centering
	\subfigbottomskip=2pt
	\subfigcapskip=-5pt
	\subfigure[]{
			\includegraphics[width=0.47\linewidth]{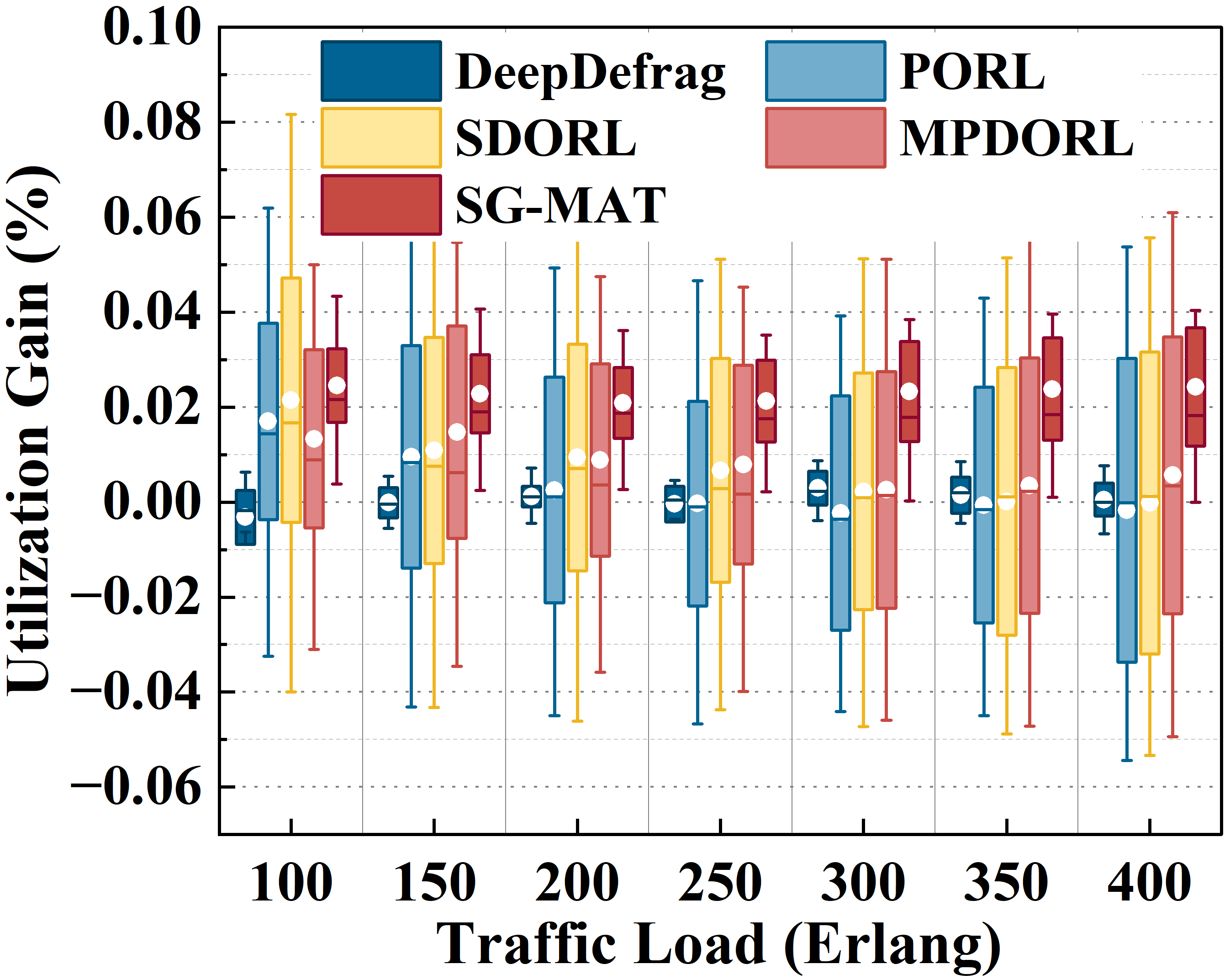}}
	\subfigure[]{
			\includegraphics[width=0.47\linewidth]{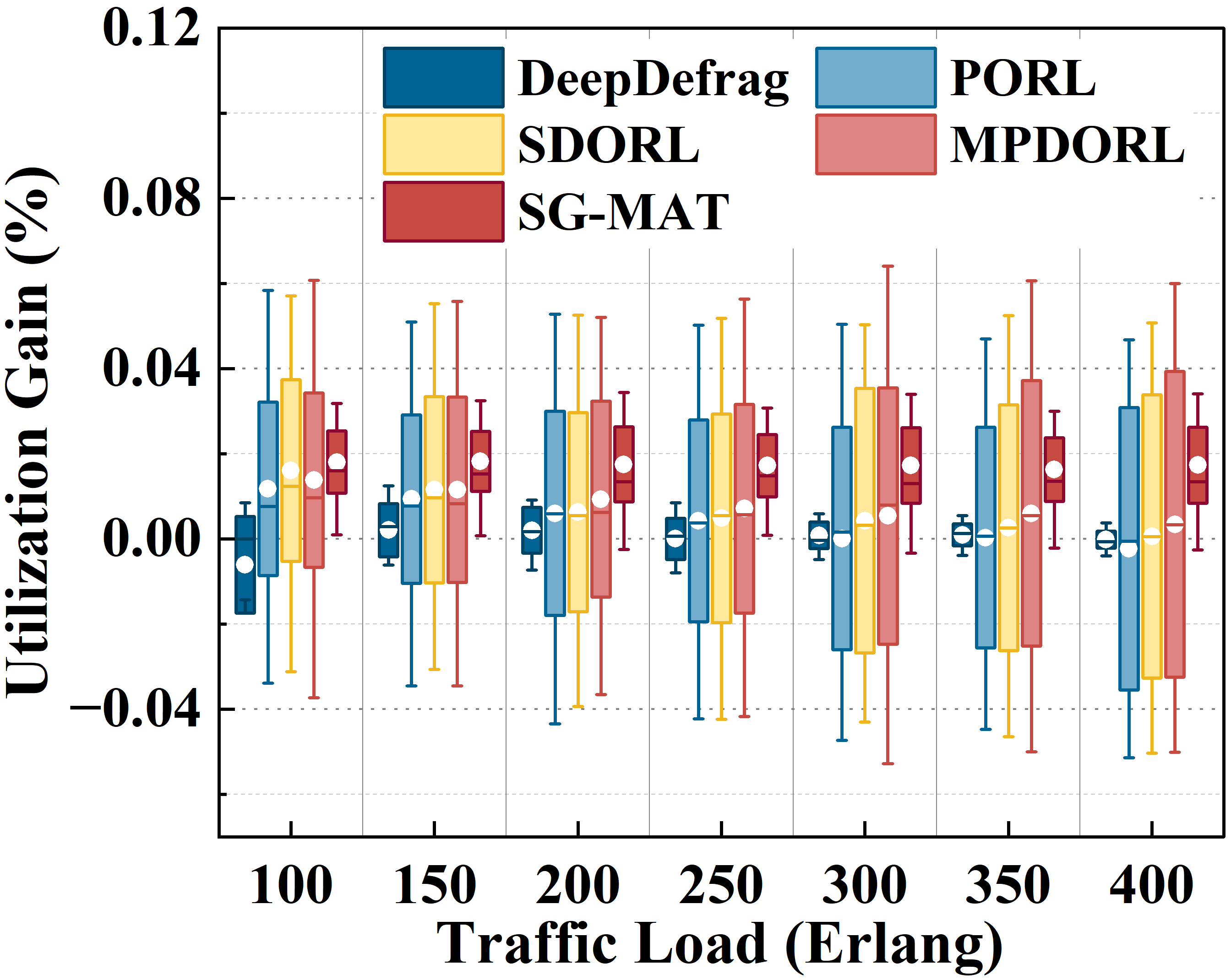}}
	\caption{\add{Average utilization gain in the system-level comparison under (a) USNET and (b) CORONET.}}
        \label{result_fig7}
\end{figure}

Fig.~\ref{result_fig7} shows the distribution of link utilization gains under varying traffic loads. SG-MAT consistently achieves the highest mean performance (white markers) with minimal variance, indicating \del{superior stability and robustness} \add{more consistent re-optimization gains}. In contrast, the baseline schemes exhibit frequent negative gains, implying that \del{certain reallocations degrade utilization due to unstable or suboptimal decisions} \add{some reallocations are less effective under the current spectrum and interference conditions}.

\begin{figure}[tp]
	\centering
	\subfigbottomskip=2pt
	\subfigcapskip=-5pt
	\subfigure[]{
			\includegraphics[width=0.47\linewidth]{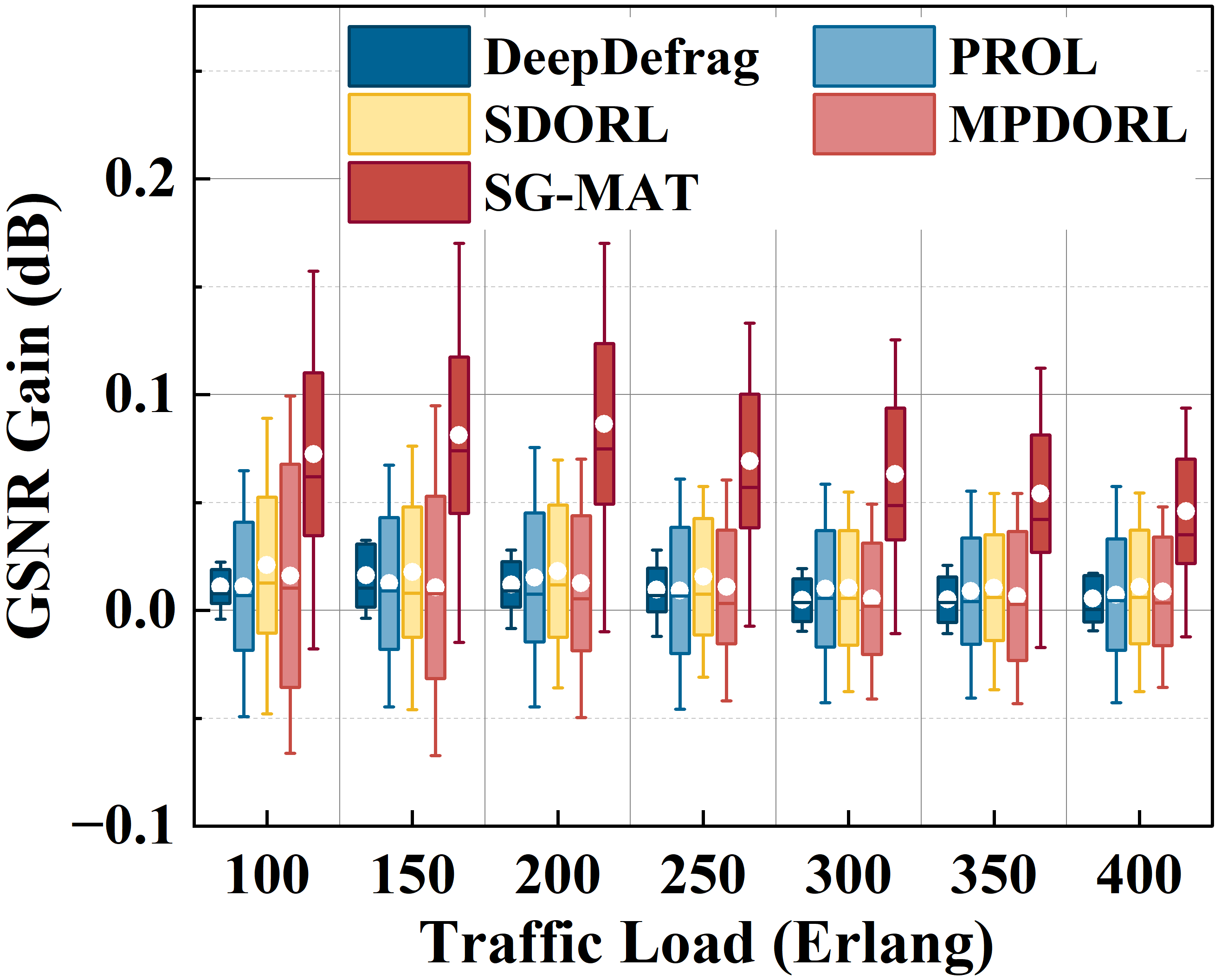}}
	\subfigure[]{
			\includegraphics[width=0.47\linewidth]{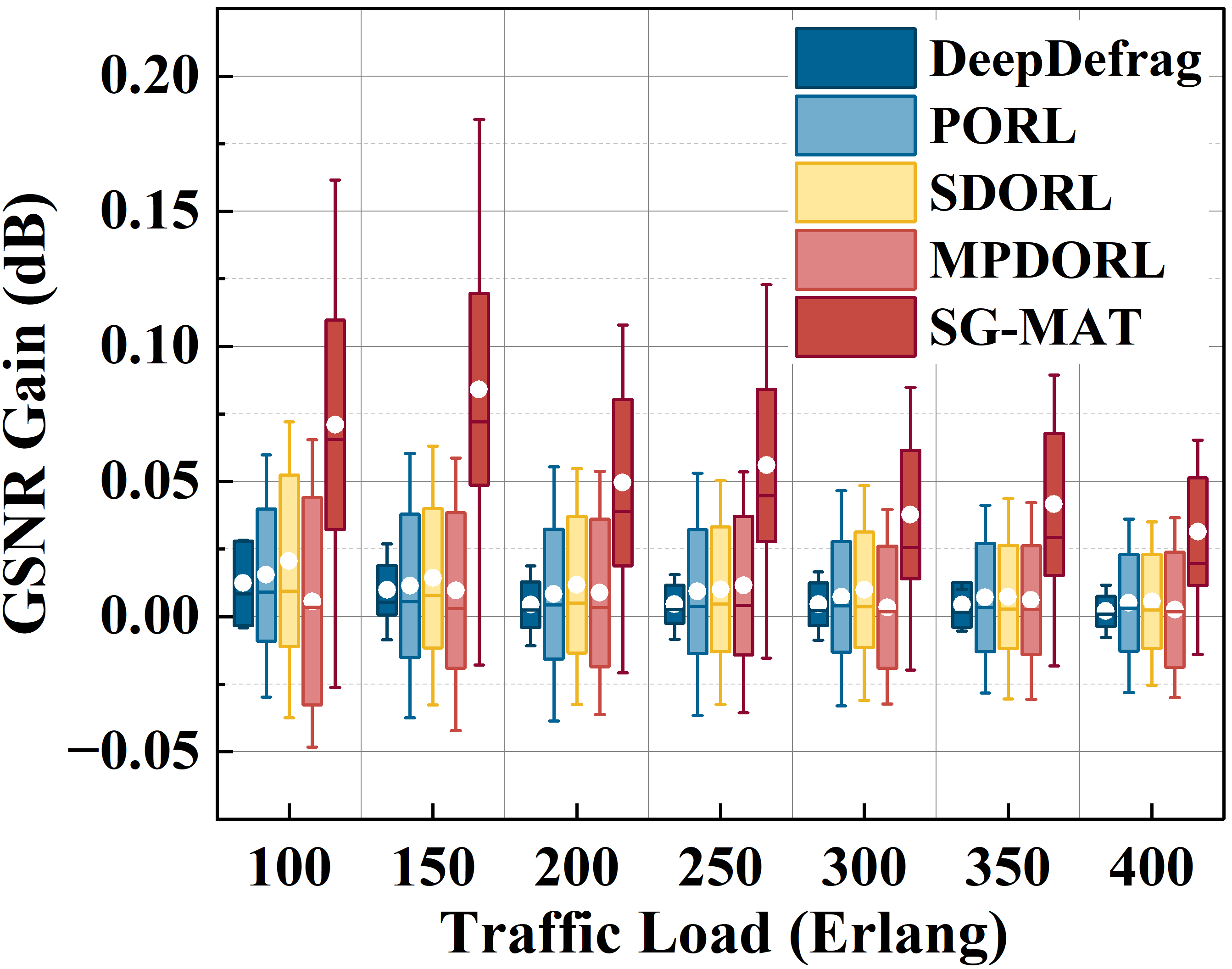}}
	\caption{\add{Average GSNR gain in the system-level comparison under (a) USNET and (b) CORONET.}}
        \label{result_fig8}
\end{figure}

Similarly, Fig.~\ref{result_fig8} presents the GSNR gain distributions. SG-MAT attains the highest average GSNR improvement across all traffic loads, with positively skewed samples that demonstrate enhanced network capacity under dynamic interference. The baselines generate noticeable negative samples, suggesting unbalanced power redistribution and nonlinear noise accumulation during re-optimization.  

These results highlight the advantage of SG-MAT’s game-theoretic coordination. Through Stackelberg-guided leader--follower interaction, SG-MAT achieves \del{stable and} consistent optimization gains. In contrast, DeepDefrag couples service selection and channel allocation into a single large action space, \del{suffering from sparse exploration and unstable convergence} \add{which may lead to inefficient exploration and limited re-optimization gains}. 
\del{PORL and SDORL lack explicit equilibrium constraints and global feedback, causing policy oscillations. Although SDORL's self-distillation mechanism alleviates policy drift through teacher--student regularization, its improvement remains limited compared with SG-MAT.}
\add{Among the ORL baselines, PORL offers online adaptability but achieves limited utilization gains; SDORL improves policy regularization via self-distillation; and MPDORL further enhances knowledge transfer through mutual policy distillation. Nevertheless, all three remain single-level policy-optimization approaches and lack explicit coordination between service selection and ISRS-aware channel reassignment, resulting in less concentrated gains than SG-MAT.}

\subsubsection{Evaluations on blocking ratio}

\begin{figure}[tp]
	\centering
	\subfigbottomskip=2pt
	\subfigcapskip=-5pt
	\subfigure[]{
			\includegraphics[width=0.47\linewidth]{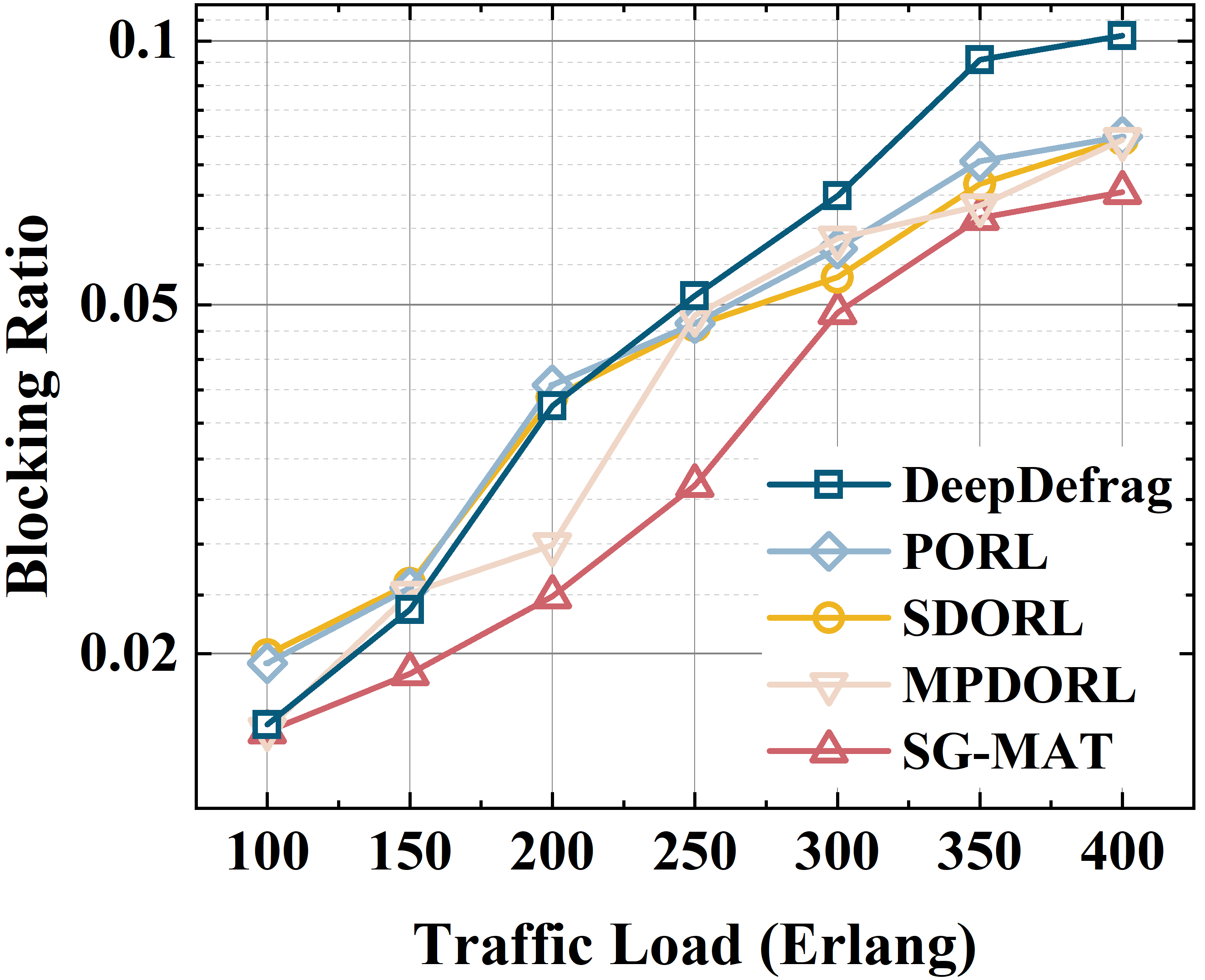}}
	\subfigure[]{
			\includegraphics[width=0.47\linewidth]{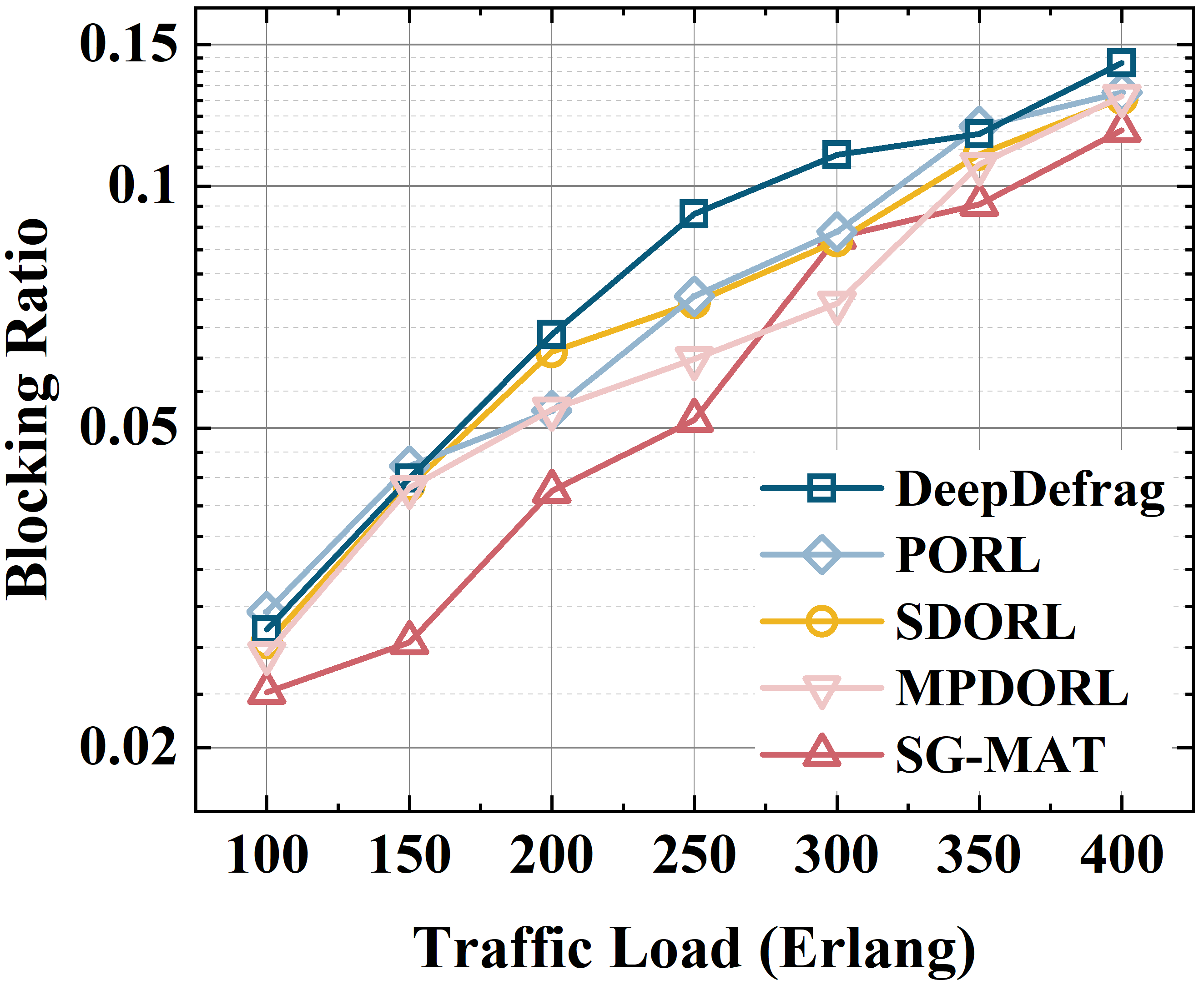}}
	\caption{\add{Blocking ratio performance in the system-level comparison under (a) USNET and (b) CORONET.}}
        \label{result_fig9}
\end{figure}

Fig.~\ref{result_fig9} depicts the blocking ratio versus traffic load. Blocking increases monotonically as spectrum resources diminish and interference coupling intensifies, and SG-MAT generally achieves stronger blocking suppression than the baselines.

In USNET, SG-MAT achieves the largest improvement in the moderate-load region (200–250~Erlangs), reducing blocking by \add{24.3\%,} 37.4\%, 38.6\%, and 39.3\% relative to \add{MPDORL,} SDORL, PORL, and DeepDefrag, respectively. This regime marks the transition from resource sufficiency to scarcity, where SG-MAT’s leader--follower coordination enables targeted reconfiguration of critical services and interference-aware resource re-optimization.

In CORONET, \add{where SG-MAT is evaluated without topology-specific retraining,} \del{SG-MAT} \add{it} also achieves \del{27–32\%} \add{26–32\%} lower blocking under light-to-moderate loads, maintaining effective re-optimization performance even as link complexity and power attenuation intensify. 
\add{This cross-topology transferability is supported by the topology-independent observations used by MAT and the event-level Stackelberg selection process, which evaluates the current candidate set using follower feedback rather than topology-specific learned patterns.}
At extreme loads, \del{differences narrow but SG-MAT still maintains a lower overall blocking ratio, confirming its effectiveness under severe congestion.}
\add{the performance gap narrows, and MPDORL even slightly outperforms SG-MAT in some cases. Nevertheless, SG-MAT maintains a lower overall blocking ratio across traffic loads, highlighting the benefit of Stackelberg-guided coordination.}

\begin{figure}[tp]
	\centering
	\subfigbottomskip=2pt
	\subfigcapskip=-5pt
	\subfigure[]{
			\includegraphics[width=0.47\linewidth]{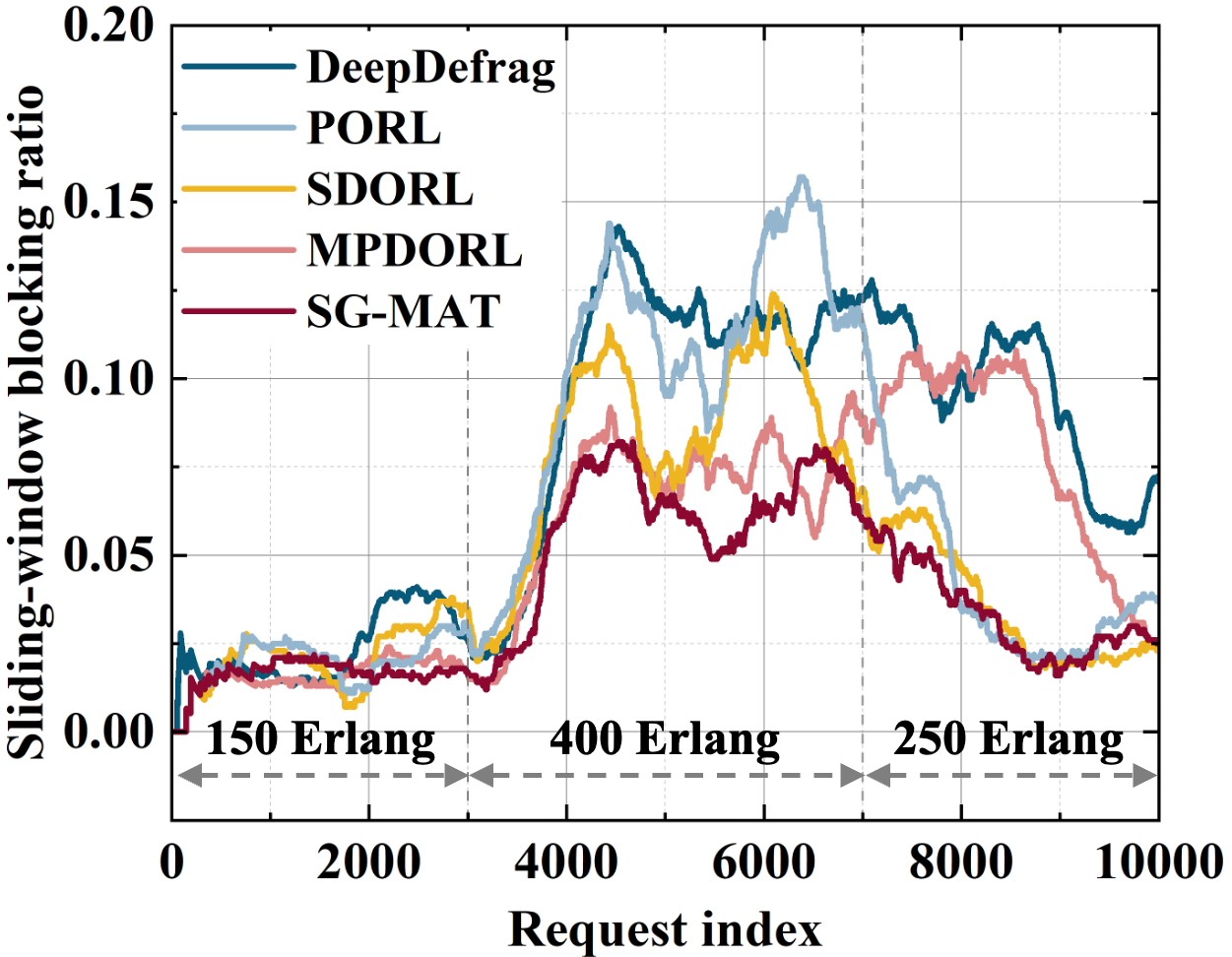}}
	\subfigure[]{
			\includegraphics[width=0.47\linewidth]{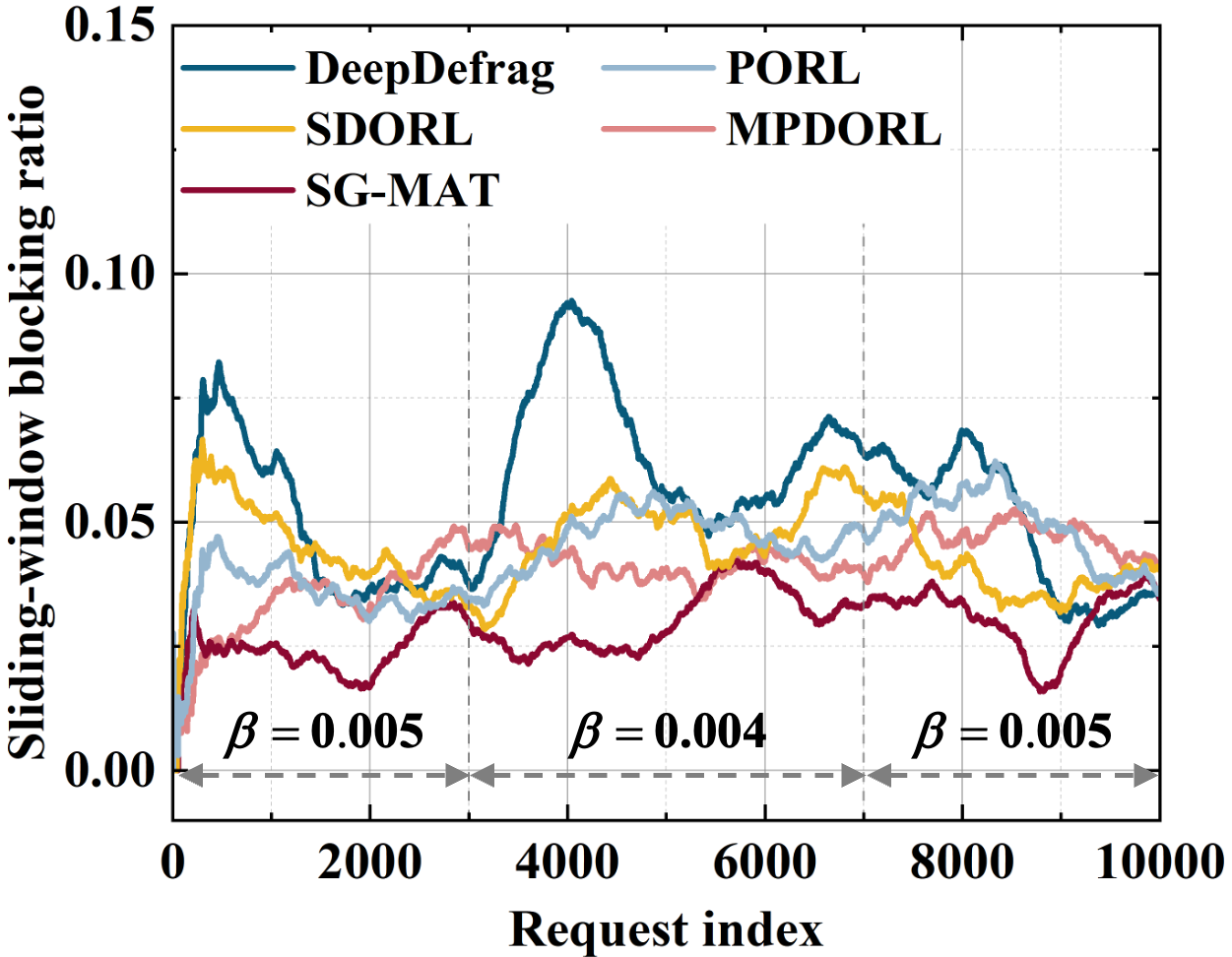}}
	\caption{\add{Sliding-window blocking ratio under (a) load-spike and (b) bitrate-mix shift scenarios on USNET.}}
        \label{result_fig14}
\end{figure}

\add{To further assess empirical robustness under traffic-pattern variations, we evaluate the local blocking ratio on USNET using a sliding window of 1000 consecutive service requests. Two scenarios are considered: a load shift from 150 to 400 Erlangs and then to 250 Erlangs, and a bitrate-mix shift at 200 Erlangs from $\beta=0.005$ to $\beta=0.004$ and back, with the intermediate phase biased toward high-bitrate services. As shown in Fig.~\ref{result_fig14}, SG-MAT consistently yields lower local blocking under both traffic shifts: its peak remains below about 0.08 during the load increase, while the baselines reach approximately 0.10--0.16; under the bitrate-mix shift, SG-MAT stays mostly below 0.04, whereas the baselines show larger fluctuations and transient peaks. These results indicate that SG-MAT provides more stable blocking suppression under the tested traffic-pattern variations.}

\subsubsection{Evaluations on re-optimization costs}

\begin{figure}[tp]
	\centering
	\subfigbottomskip=2pt
	\subfigcapskip=-5pt
	\subfigure[]{
			\includegraphics[width=0.47\linewidth]{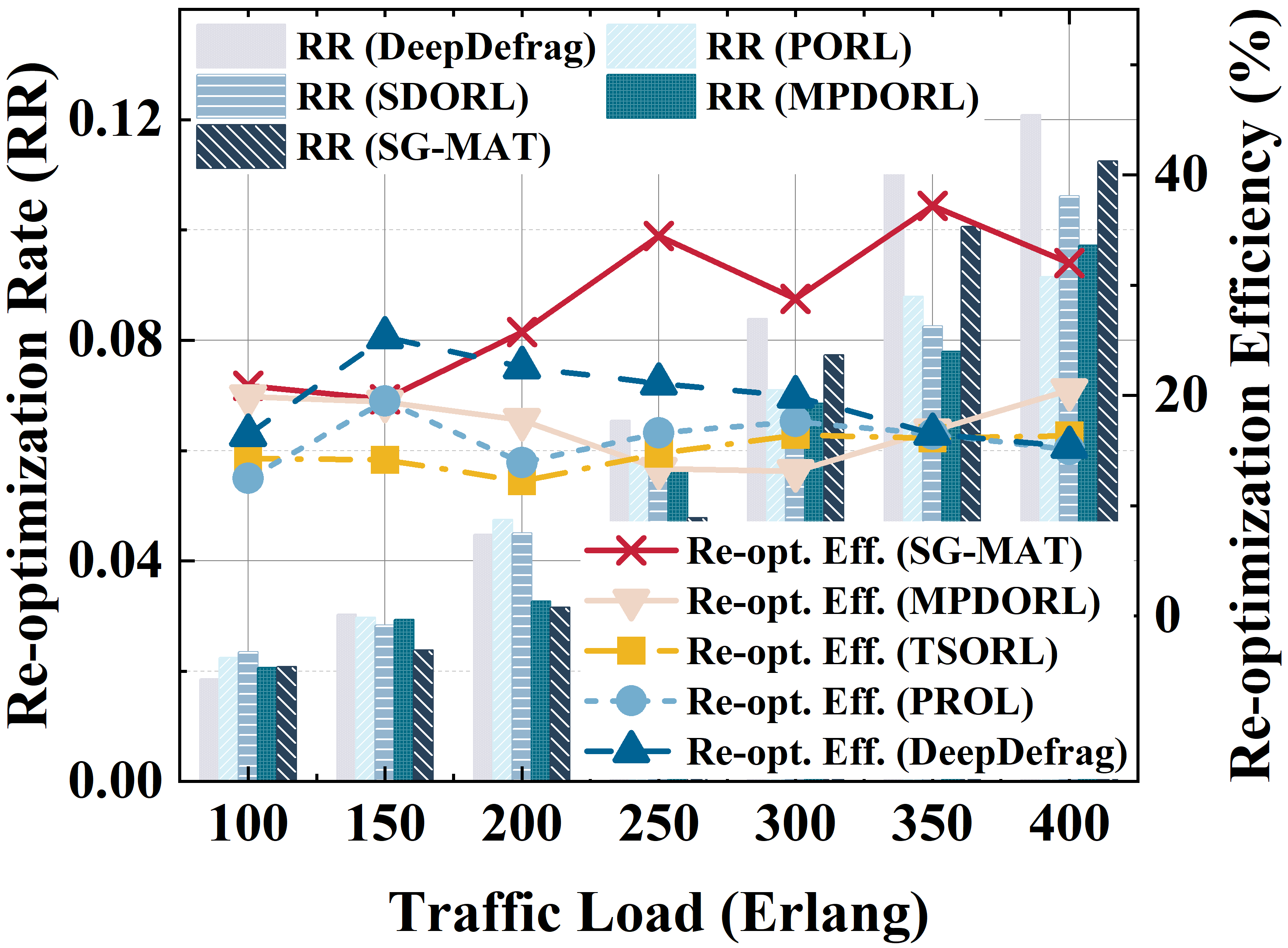}}
	\subfigure[]{
			\includegraphics[width=0.47\linewidth]{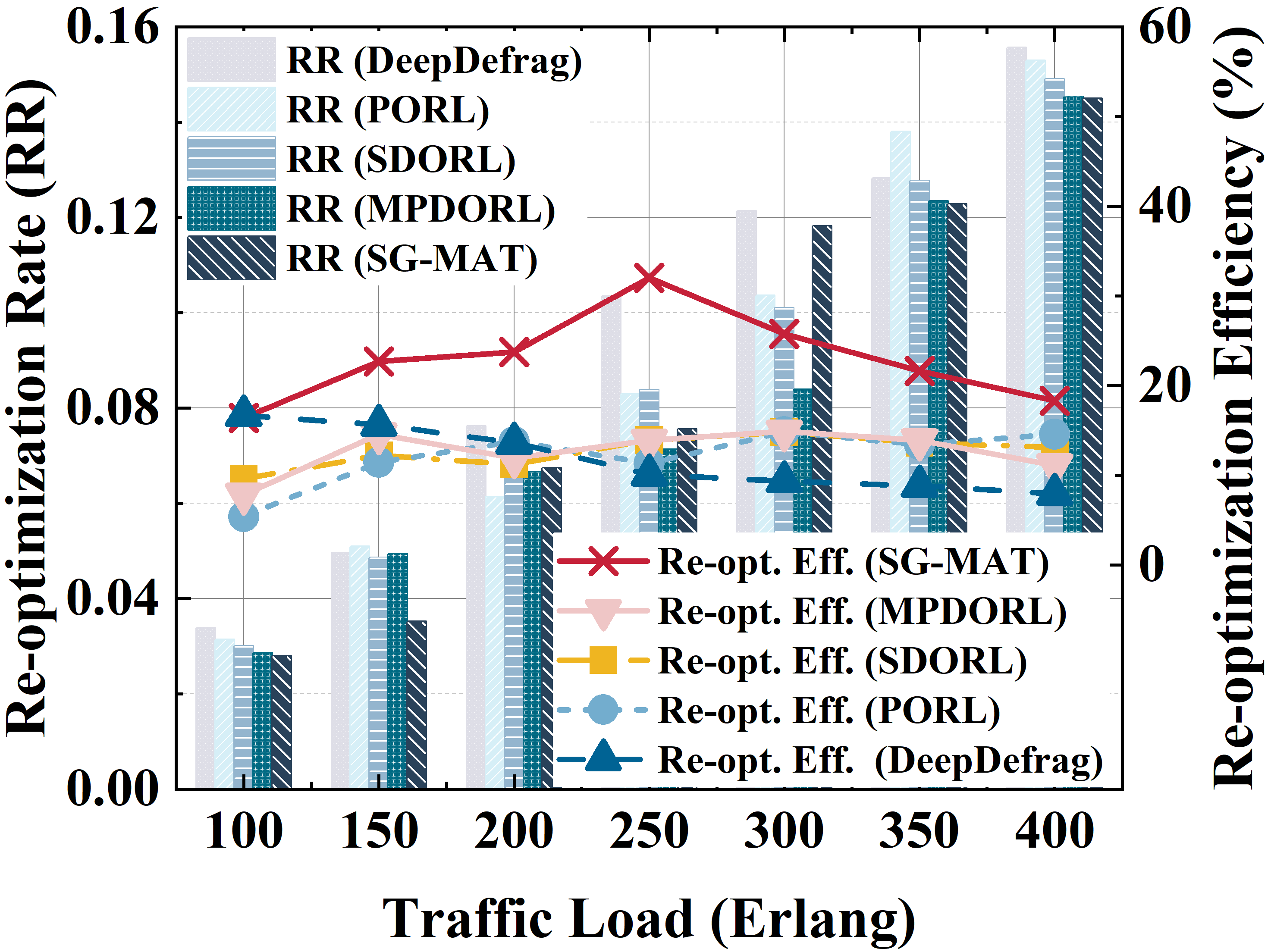}}
	\caption{\add{Re-optimization frequency and efficiency in the system-level comparison under (a) USNET and (b) CORONET.}}
	\label{result_fig10}
\end{figure}

Fig.~\ref{result_fig10} compares the evolution of re-optimization rate and efficiency. All algorithms perform a larger number of re-optimizations as the traffic load increases; however, SG-MAT maintains the highest efficiency while keeping the number of operations moderate. Although its re-optimization count slightly exceeds \del{PORL and SDORL} \add{PORL, SDORL, and MPDORL} under heavy load in USNET, the per-operation gain is substantially higher, indicating a superior benefit--cost balance.  
DeepDefrag triggers the most frequent but least effective adjustments due to its flat, unselective action space.

\begin{figure}[tp]
	\centering
	\subfigbottomskip=2pt
	\subfigcapskip=-5pt
	\subfigure[]{
			\includegraphics[width=0.47\linewidth]{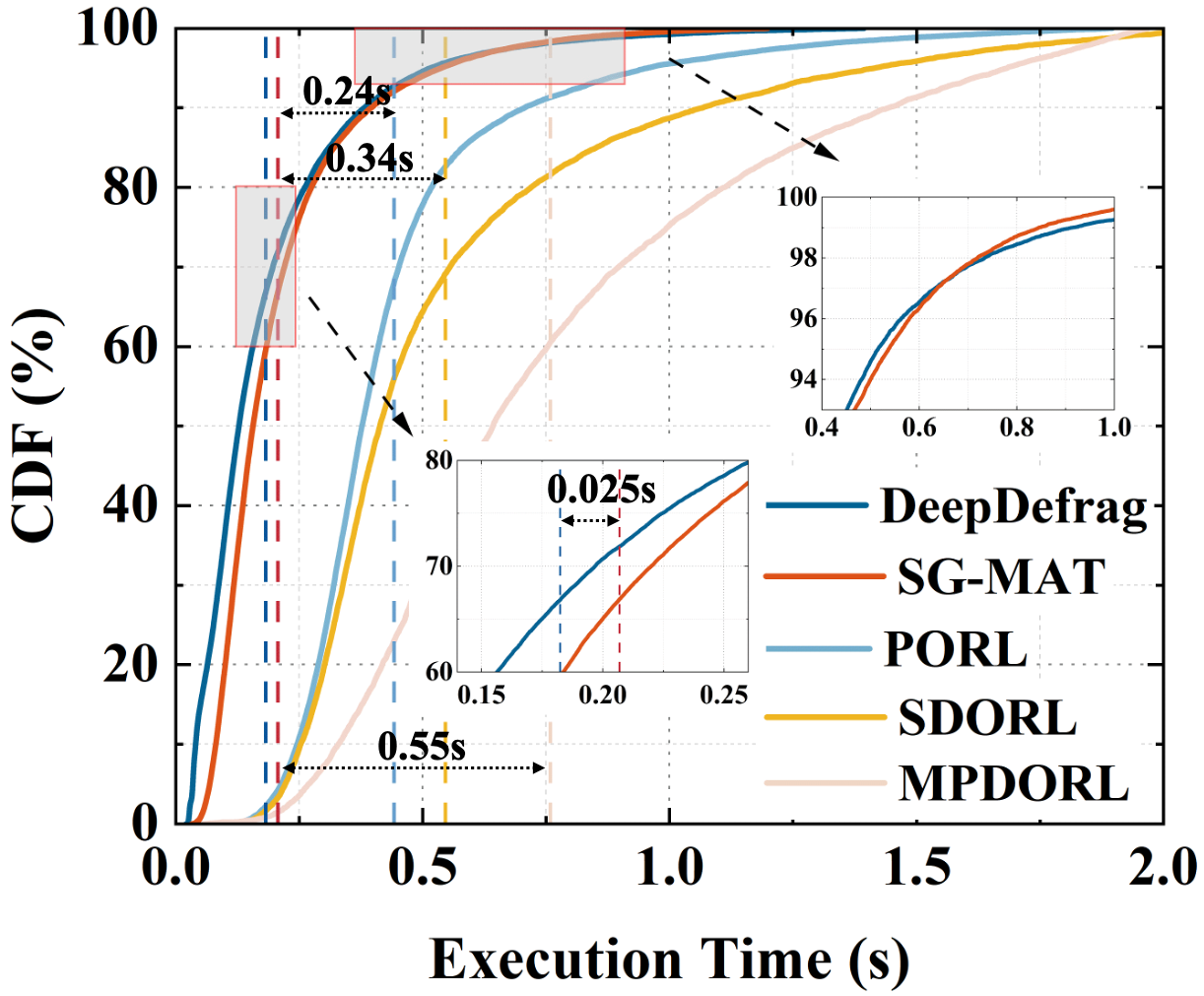}}
	\subfigure[]{
			\includegraphics[width=0.47\linewidth]{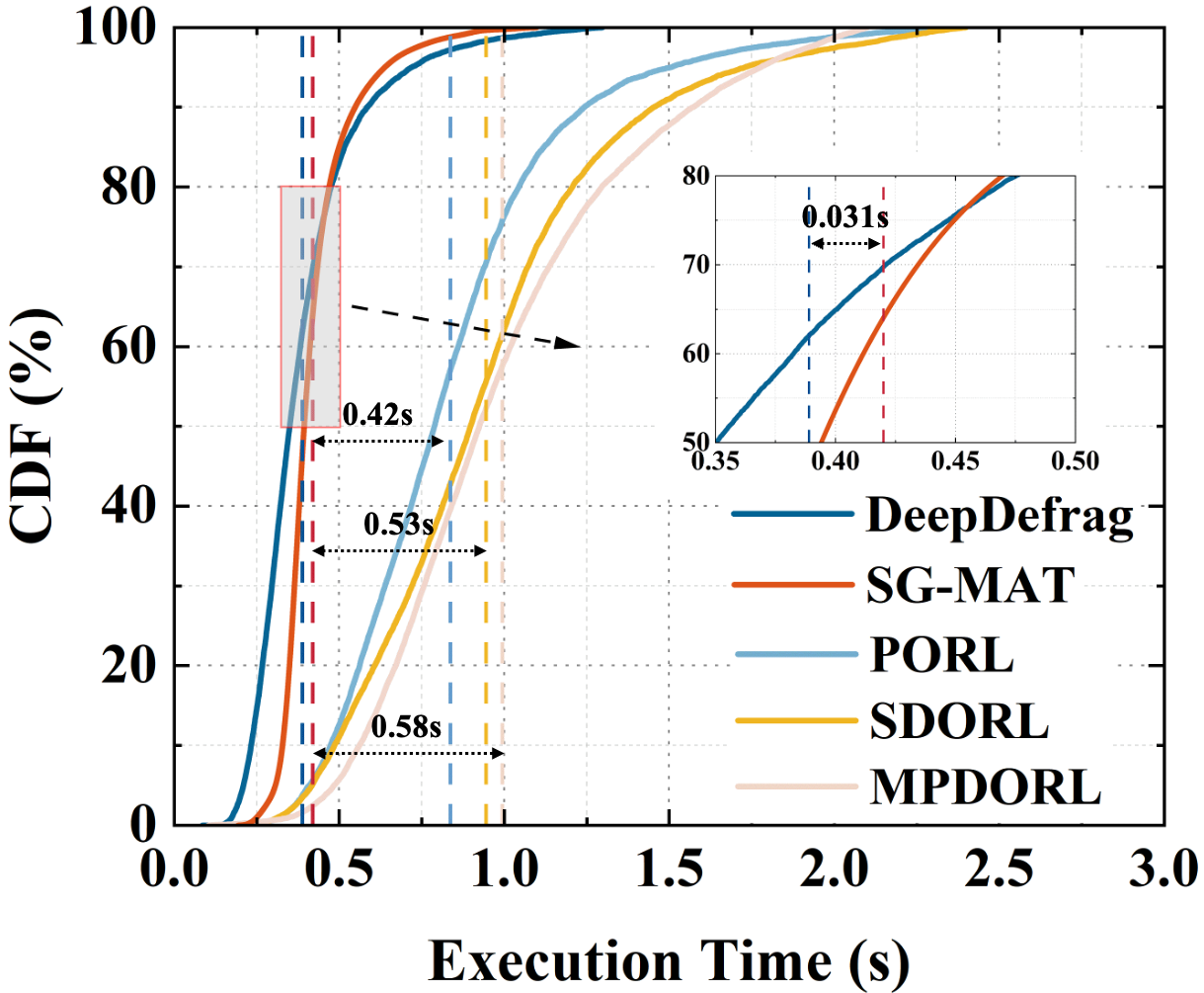}}
	\caption{\add{CDF of the re-optimization execution time in the system-level comparison under (a) USNET and (b) CORONET.}}
	\label{result_fig11}
\end{figure}

The CDF of execution time in Fig.~\ref{result_fig11} further demonstrates SG-MAT's computational efficiency. While its average latency exceeds DeepDefrag by only 25~ms in USNET and 31~ms in CORONET, SG-MAT exhibits tighter temporal dispersion and fewer long-tail outliers, benefiting from improved decision quality and reduced feasibility-verification overhead. In contrast, PORL incurs additional delay from online policy updates, and SDORL suffers additional overhead from its self-distillation alignment process. \add{MPDORL introduces additional knowledge-transfer and mutual-distillation operations, resulting in higher runtime overhead.}

\add{The re-optimization cost results further support the online deployment feasibility of SG-MAT. With millisecond-level decision latency, a centralized controller can promptly obtain the service subset and channel reassignment upon a blocking event and translate them into device-level commands for reconfigurable optical add-drop multiplexers (ROADMs) and flexible transponders~\cite{N. Sambo}. By targeting critical services, SG-MAT also reduces ineffective reconfiguration attempts, control-plane signaling, and service disruption risk.}

Overall, SG-MAT outperforms all baselines in utilization and GSNR improvement, blocking suppression, and re-optimization efficiency while maintaining competitive runtime. The results suggest that the Stackelberg game–driven architecture \del{effectively achieves dynamic equilibrium between global coordination and adaptation} \add{enables feedback-guided coordination between service selection and ISRS-aware channel reassignment}. By integrating equilibrium guidance with multi-agent modeling, SG-MAT \del{ensures precise, stable, and scalable optimization} \add{achieves accurate, consistent, and scalable re-optimization} for multi-band optical networks.

\subsection{\add{Ablation Study of SG-MAT}}
\add{
Sections VI-B and VI-C have validated the effectiveness of the MAT follower and the SG-MAT framework, respectively. This subsection further presents a component-wise ablation study to quantify the contributions of two key modules: the MAT-based channel reassignment layer and the Stackelberg game-based service selection layer.}

\begin{figure}[tp]
	\centering
	\subfigbottomskip=2pt
	\subfigcapskip=-5pt
	\subfigure[]{
			\includegraphics[width=0.47\linewidth]{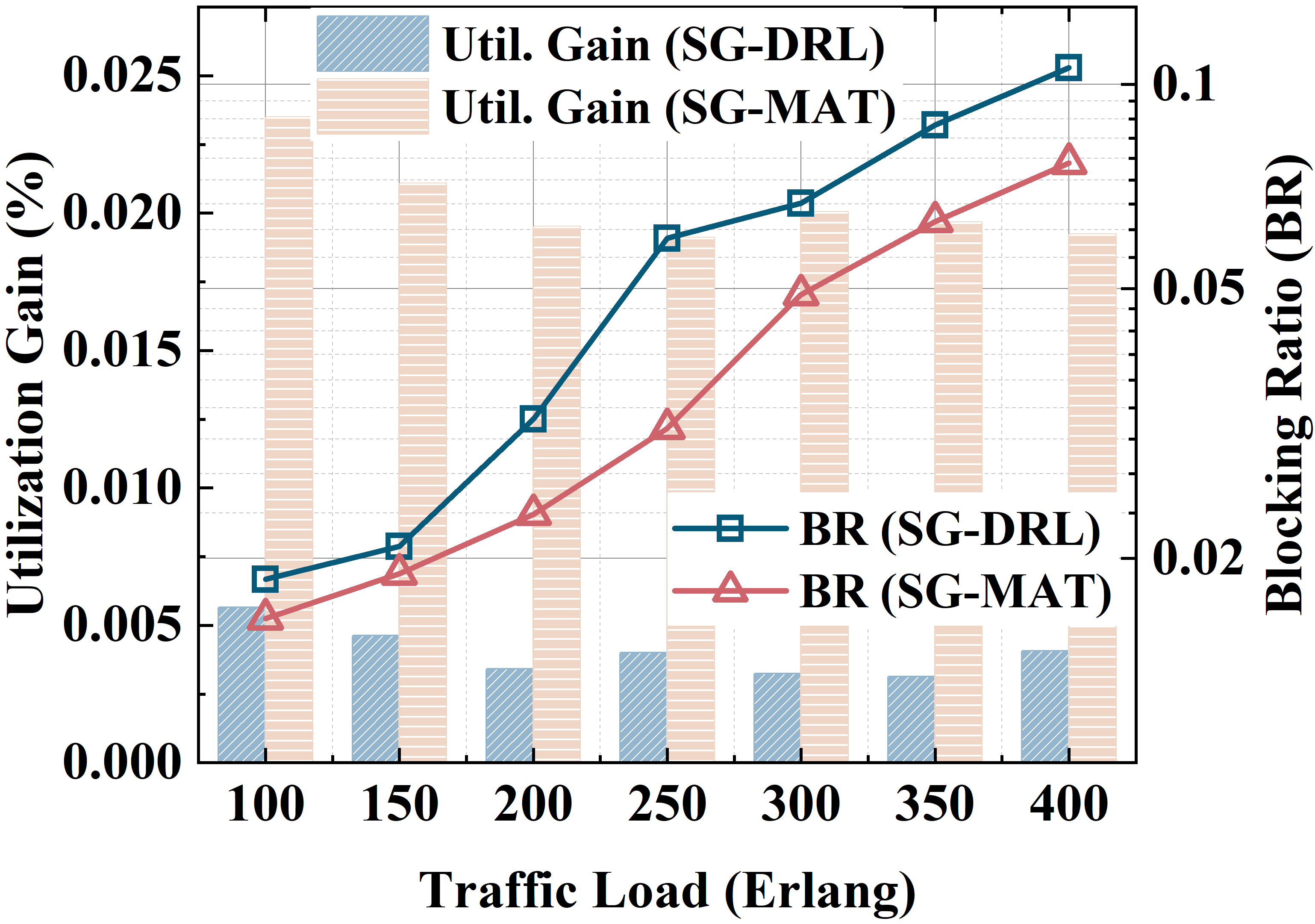}}
	\subfigure[]{
			\includegraphics[width=0.47\linewidth]{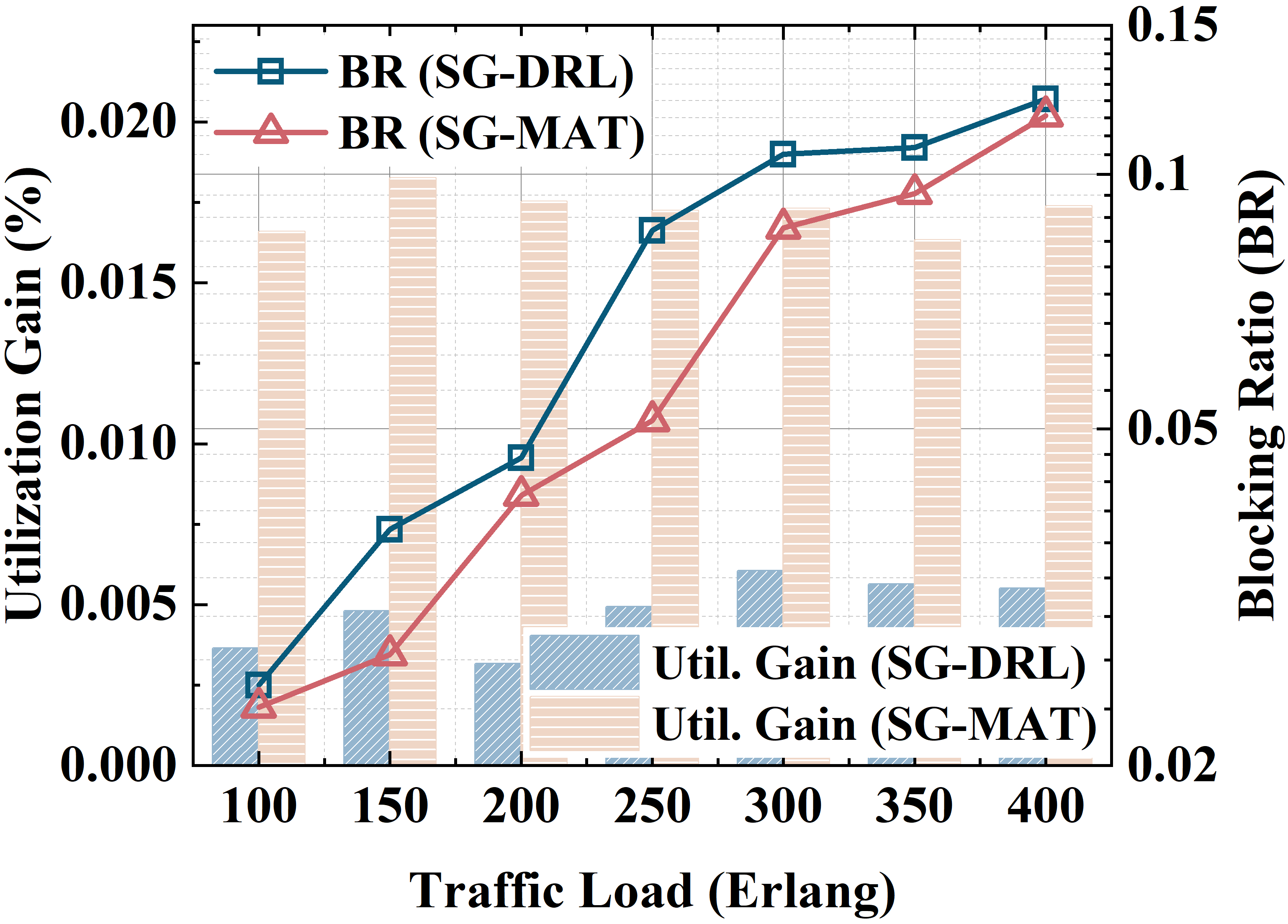}}
	\caption{\add{Utilization gain and blocking ratio of SG-DRL and SG-MAT under (a) USNET and (b) CORONET.}}
	\label{result_fig12}
\end{figure}

\add{
To evaluate the contribution of inter-service attention modeling, we construct SG-DRL as an ablation variant. SG-DRL retains the same Stackelberg selection layer as SG-MAT but replaces the MAT follower with the DRL baseline in Section VI-B. As shown in Fig.~\ref{result_fig12}, SG-MAT achieves higher utilization gains and lower blocking ratios than SG-DRL across both topologies and different traffic loads. This indicates that MAT can better capture nonlinear inter-service coupling and generate more coordinated multi-service reconfiguration decisions.
}

\begin{figure}[tp]
	\centering
	\subfigbottomskip=2pt
	\subfigcapskip=-5pt
	\subfigure[]{
			\includegraphics[width=0.47\linewidth]{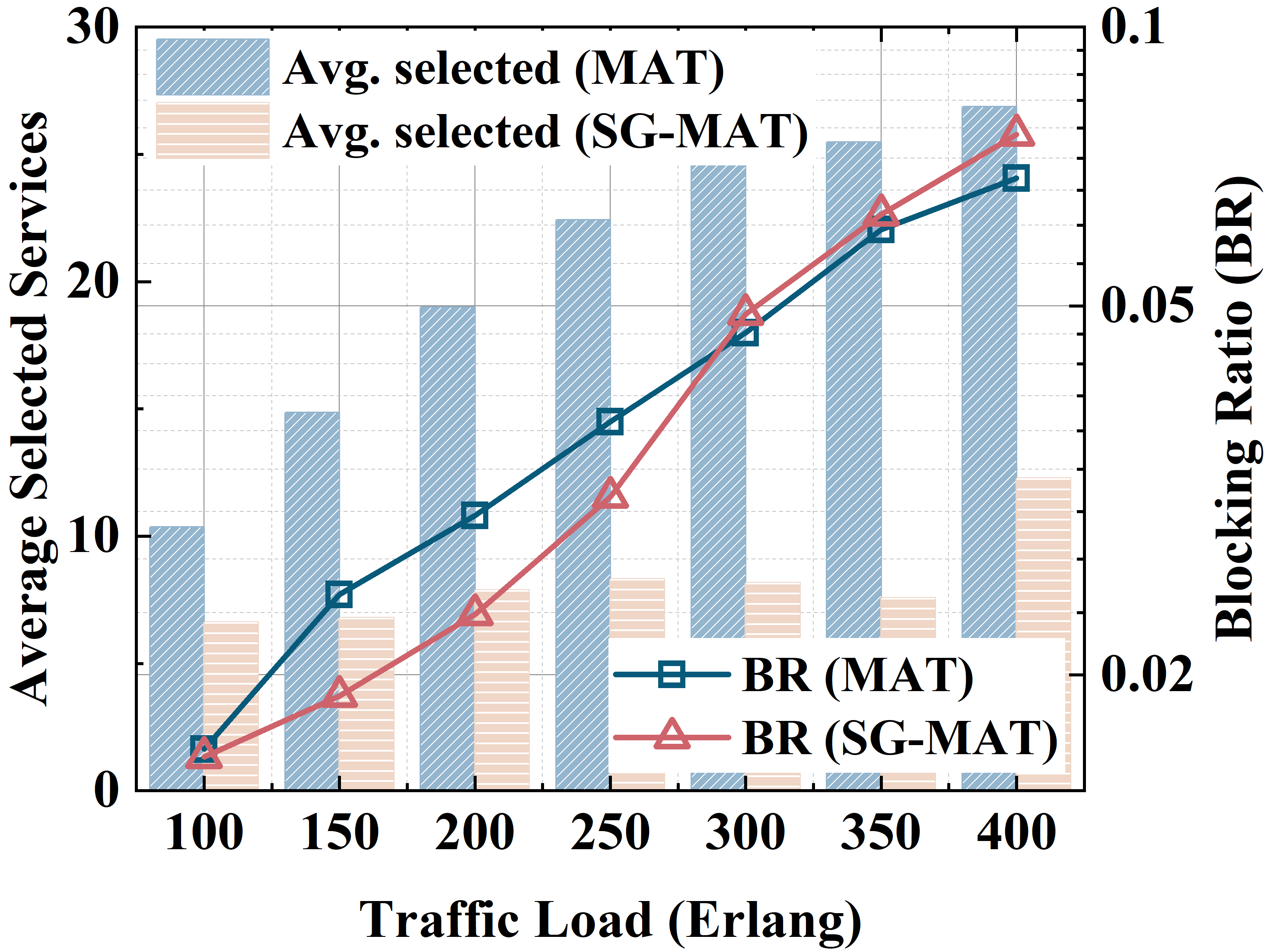}}
	\subfigure[]{
			\includegraphics[width=0.47\linewidth]{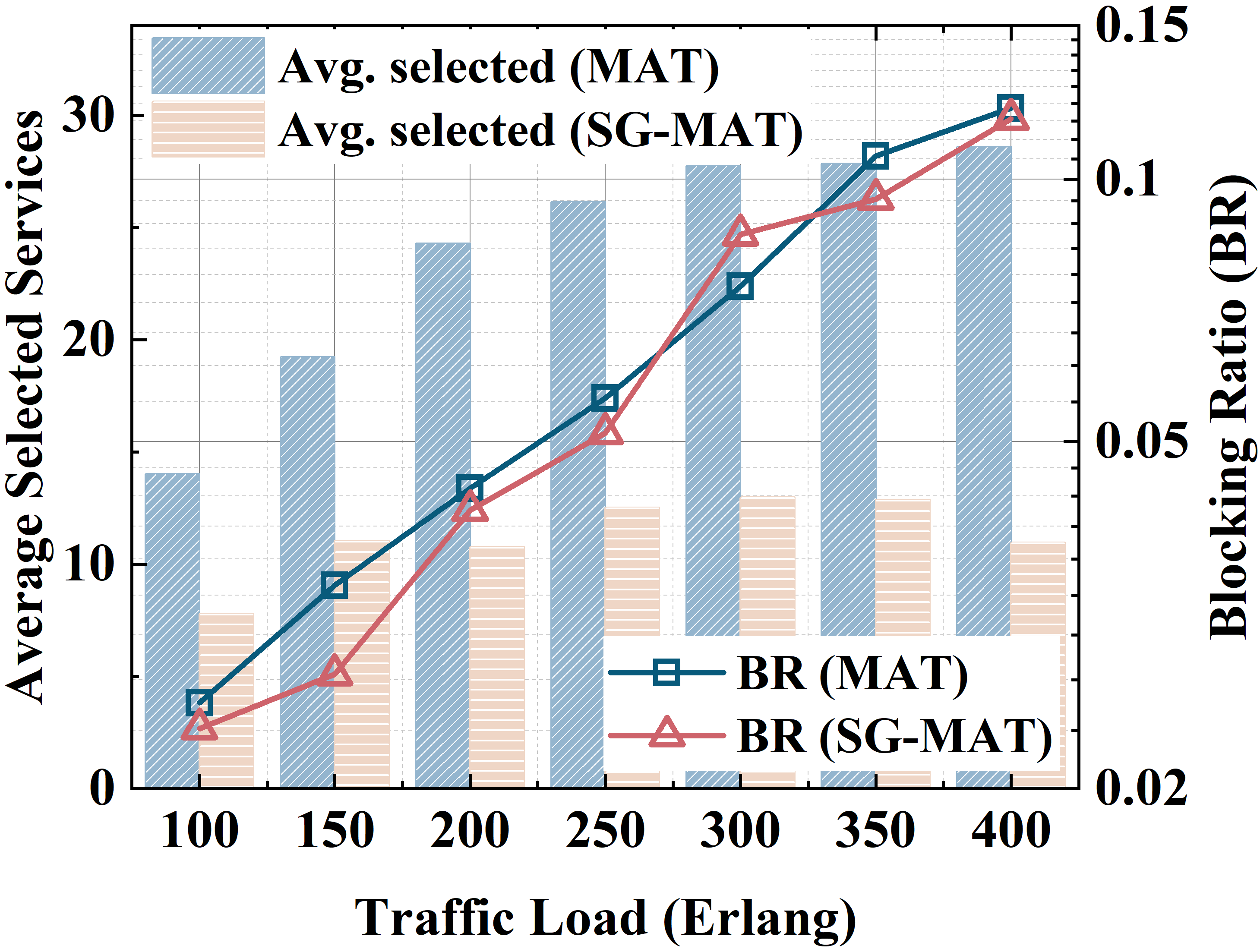}}
	\caption{\add{Average selected services and blocking ratio of MAT and SG-MAT under (a) USNET and (b) CORONET.}}
	\label{result_fig13}
\end{figure}

\add{
To assess the contribution of the Stackelberg selection layer, we use MAT in Section VI-B as the ablation variant, where services are selected by a rule-based mechanism. Fig.~\ref{result_fig13} shows that MAT selects more services for re-optimization, whereas SG-MAT achieves lower or comparable blocking ratios with fewer selected services. This indicates that the Stackelberg layer improves selection quality and makes re-optimization more targeted and efficient.
}

\add{
Overall, these ablation results show that the performance gains of SG-MAT stem from the synergy between its two core modules: the Stackelberg game layer improves the selection quality of the re-optimized service set, while MAT enhances the coordinated adjustment capability for a given service set.
}

\section{Conclusion}

In this paper, we investigated the problem of ISRS-aware resource re-optimization in multi-band optical networks. We formulated the re-optimization process as a leader–follower Stackelberg game to establish a structured decision framework linking global service subset selection with coordinated multi-service reallocation. At the follower layer, we introduced MAT to characterize ISRS-induced nonlinear inter-service coupling and generate coordinated channel-adjustment strategies. At the leader layer, we proved the existence of SE and showed that BI characterizes the equilibrium solution of the finite Stackelberg game. Building on this theoretical foundation, we developed GLSE, which combines greedy construction and local refinement to obtain an equilibrium-guided leader decision with polynomial complexity. Simulation results demonstrate that SG-MAT achieves higher utilization improvement and re-optimization efficiency than existing ORL baselines, and reduces the blocking ratio by approximately 30\% under moderate traffic loads on both USNET and CORONET topologies.
\add{A promising future direction is to extend SG-MAT from the single-administrative-domain setting to multi-administrative-domain inter-AIDC networks, where coupled re-optimization decisions under limited information sharing could be explored through federated multi-leader--multi-follower Stackelberg formulations.}

\end{document}